\documentclass[%
 aip,
 amsmath,amssymb,
 reprint,%
]{revtex4-1}

\usepackage{graphicx}
\usepackage{dcolumn}
\usepackage{bm}

\usepackage[utf8]{inputenc}
\usepackage[T1]{fontenc}
\usepackage{mathptmx}
\usepackage{etoolbox}
\usepackage{hyperref}

\makeatletter
\def\@email#1#2{%
 \endgroup
 \patchcmd{\titleblock@produce}
  {\frontmatter@RRAPformat}
  {\frontmatter@RRAPformat{\produce@RRAP{*#1\href{mailto:#2}{#2}}}\frontmatter@RRAPformat}
  {}{}
}%
\makeatother
\begin{document}

\preprint{AIP/123-QED}

\title[Wafer-scale correlated morphology and optoelectronic properties in GaAs/AlGaAs core-shell nanowires]{Wafer-scale correlated morphology and optoelectronic properties in GaAs/AlGaAs core-shell nanowires}
\author{Ishika Das}

 \email{ishika.das-2@postgrad.manchester.ac.uk}
 
\affiliation{ 
Department of Physics and Astronomy and the Photon Science Institute, University of Manchester, Manchester, M13 9PL, United Kingdom
}%

\author{Keisuke Minehisa}
\author{Fumitaro Ishikawa}
 
\affiliation{%
Research Center for Integrated Quantum Electronics, Hokkaido University, Sapporo, Japan%
}%

\author{Patrick Parkinson}
\author{Stephen Church}
\email{stephen.church@manchester.ac.uk}

\affiliation{ 
Department of Physics and Astronomy and the Photon Science Institute, University of Manchester, Manchester, M13 9PL, United Kingdom
}%

\date{\today}

\begin{abstract}
Achieving uniform nanowire size, density, and alignment across a wafer is challenging, as small variations in growth parameters can impact performance in energy harvesting devices like solar cells and photodetectors. This study demonstrates the in-depth characterization of uniformly grown GaAs/AlGaAs core-shell nanowires on a two-inch Si(111) substrate using Ga-induced self-catalyzed molecular beam epitaxy. By integrating Scanning Electron Microscopy and Time Correlated Single-Photon Counting, we establish a detailed model of structural and optoelectronic properties across wafer and micron scales. While emission intensity varies by up to 35\%, carrier lifetime shows only 9\% variation, indicating stable material quality despite structural inhomogeneities. These findings indicate that, for the two-inch GaAs/AlGaAs nanowire wafer, achieving uniform nanowire coverage had a greater impact on consistent optoelectronic properties than variations in material quality, highlighting its significance for scalable III-V semiconductor integration on silicon in advanced optoelectronic devices such as solar cells and photodetectors.
\end{abstract}

\maketitle

\begin{figure*}
\includegraphics[width=1\linewidth]{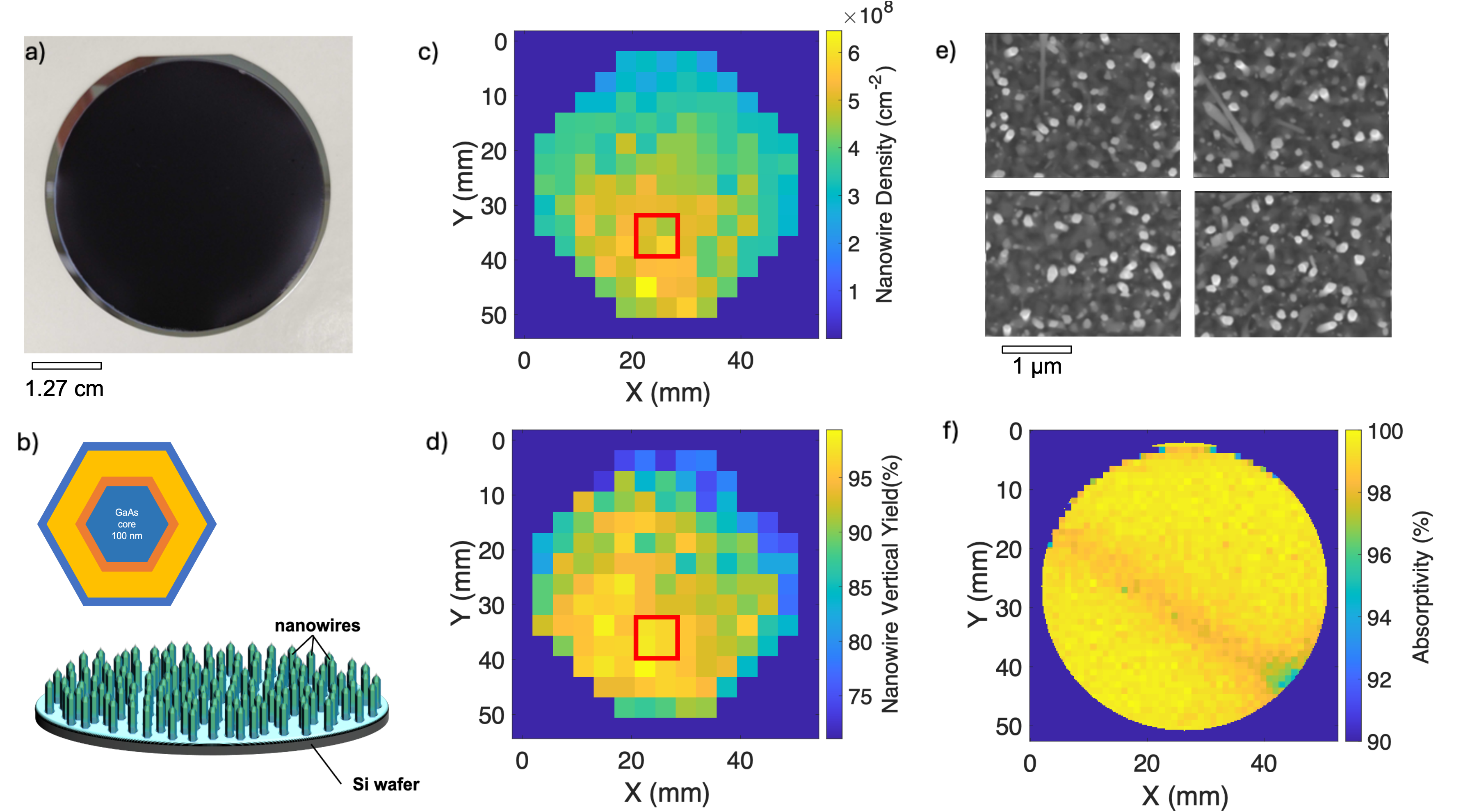}
\caption{\label{nanowire density}(a) Two-inch wafer containing GaAs/AlGaAs core-shell nanowires, with a mostly uniform matte-black appearance indicating coverage of nanowires. (b)  Cross-sectional view of a single nanowire, featuring a 100 nm GaAs core (blue), a 5 nm AlGaAs shell (orange), a 75 nm p-doped AlGaAs layer (yellow), and a 5 nm p$^{+}$ doped GaAs outer layer (blue); and a schematic illustration of the two-inch wafer showing vertical alignment of nanowires. (c, d) Maps displaying nanowire density and vertical yield across the two-inch wafer, with (e) providing SEM images of the areas highlighted in red on the maps in (c) and (d). (f) Absorptivity map derived from reflectivity measured across the wafer.}
\end{figure*}

\begin{figure*}
\includegraphics[width=1\linewidth]{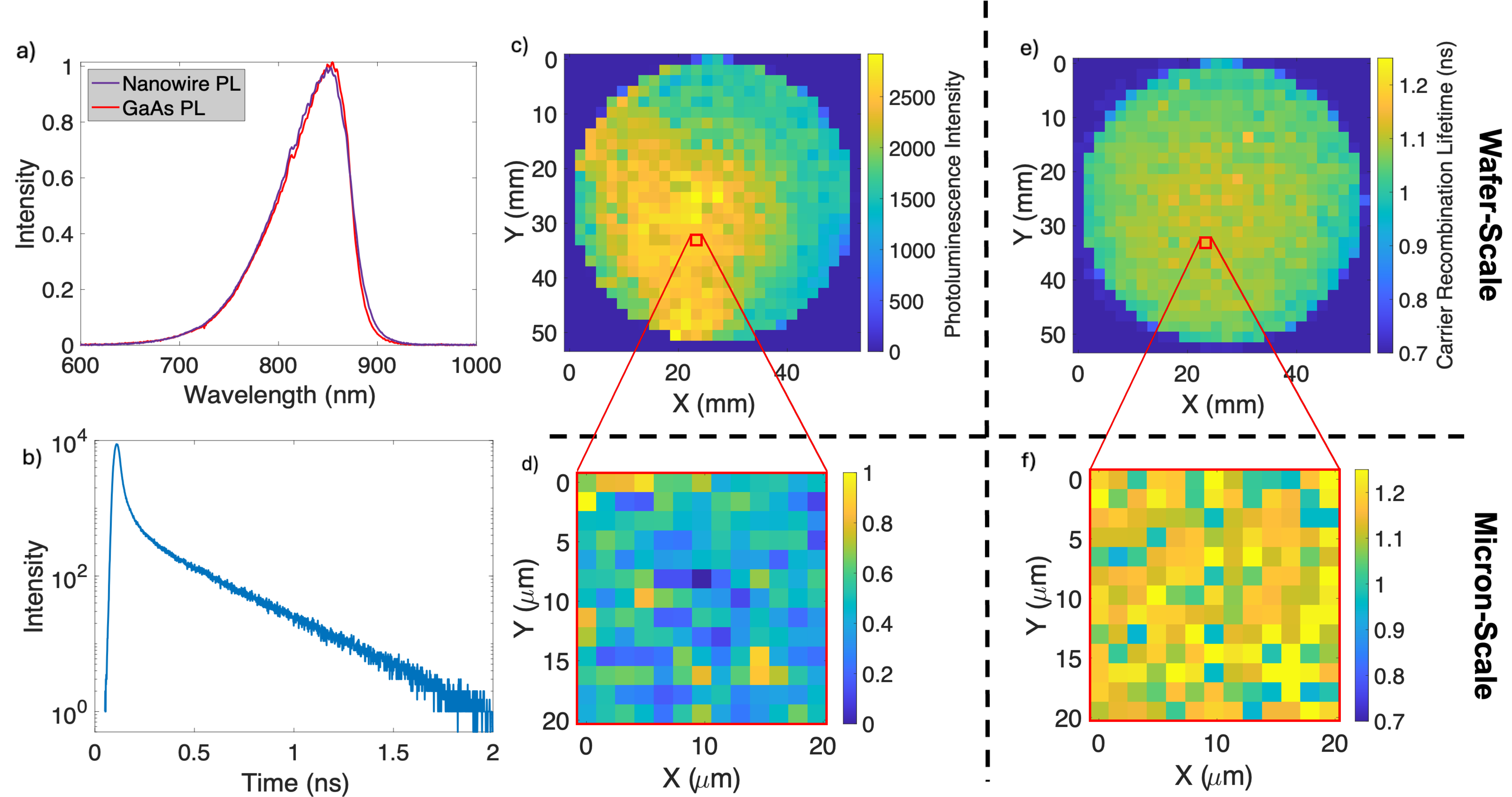}
\caption{\label{TCSPC map}(a) PL spectrum of a reference GaAs wafer (red) compared to GaAs/AlGaAs core-shell nanowires (purple). (b) Median carrier recombination decay plot of the nanowires measured using time-correlated single photon counting (TCSPC). (c) Wafer-scale PL intensity map of the nanowires. (d) Local-scale PL intensity variation map, shown in red box of the wafer-scale map in (c). (e) Wafer-scale carrier lifetime map of the nanowires. (f) Local-scale carrier lifetime variation map, shown in red box of the wafer-scale map in (e). In (c) and (e), each pixel represents the standard deviation from the median values plotted in (d) and (f), respectively.}
\end{figure*}

\begin{figure*}
\includegraphics[width=1\linewidth]{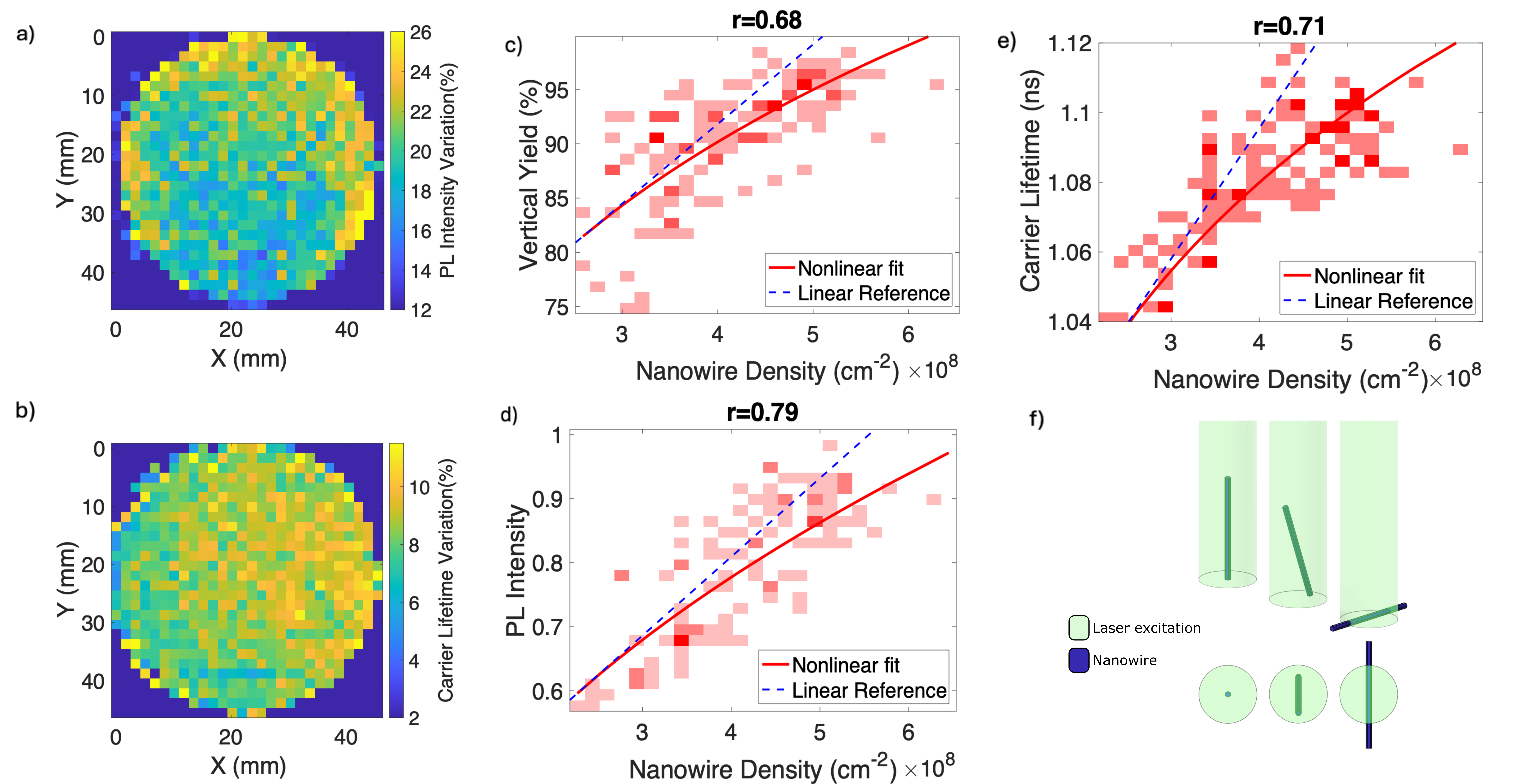}
\caption{\label{Variations}(a) Micron-scale variation in PL intensity, showing a local standard deviation of 21\% compared to a 35\% variation at the wafer scale. 
(b) Micron-scale variation in carrier recombination lifetime, with both local and wafer-scale variations exhibiting a consistent 9\% standard deviation from the median. 
(c-e) 2D histogram plots between nanowire density and (c) vertical yield, (d) PL intensity, and (e) carrier recombination lifetime. The red squares represent 2D bins, with the colour saturation indicating the number of datapoints in each bin. \(r\) represents the Pearson correlation coefficient.
All plots reveal a sublinear increase with nanowire density (red solid line), with blue dotted lines indicating a reference linear relationship. (f) Schematic showing how the area which laser excites changes with nanowires aligned from vertical to horizontal.
}
\end{figure*}


Nanowires offer a way to integrate optoelectronic materials, such as III-V semiconductors, with electronic materials, like silicon\cite{Martensson2004EpitaxialSilicon}. This integration is challenging due to lattice mismatch, but at the nanoscale, strain relief provides a way to overcome this\cite{Chuang2007CriticalSubstrates}\textsuperscript{,}\cite{Borg2014VerticalSi100}, enabling the growth of high quality III-V material on silicon, which not only takes advantage of the mature silicon industry and existing foundries, but also enhances the potential for electronic and photonic integration on a single platform\cite{Goktas2018NanowiresReview}. Furthermore, the light absorption properties of these structures is often enhanced relative to their conventional material properties due to enhanced light-scattering\cite{Kelzenberg2010EnhancedApplications}\textsuperscript{,}\cite{Muskens2008DesignApplications}, which is advantageous for light-harvesting applications\cite{Diedenhofen2011StrongNanowires}. Epitaxial nanowire growth methods such as molecular beam epitaxy (MBE) allow for precise control over nanowire morphology, structure, and composition, facilitating the formation of heterostructures with tailored electronic and optical properties essential for optoelectronics and quantum technologies\cite{Mata2013ANanowire}\textsuperscript{,} \cite{Gudiksen2002GrowthElectronics}. 

Core-shell nanowires, consisting of a central core material surrounded by concentric layers (shells) of different materials, passivate the surface and reduce nonradiative surface recombination\cite{Chen2015SuppressionNanowires}, which enhances the efficiency of light-emitting and detecting devices\cite{Chen2011NanolasersSilicon,Paramasivam2023Self-consistentHeterostructure}. The heterostructure also facilitates tuneability over electronic and photonic properties, enabling the development of advanced devices with tailored performance characteristics\cite{Lauhon2002EpitaxialHeterostructures}. Self-catalyzed growth allows scalable production with reduced impurity risks\cite{Glas2013PredictiveGrowth} and also  demonstrates successful integration with various substrates and high yields, particularly for materials such as GaAs and InAs\cite{Barrigon2019SynthesisNanowires}. Using this with MBE will allow precise control over growth conditions and material quality for large-scale integration of semiconductor nanowires\cite{Jabeen2008Self-catalyzedEpitaxy}. However, achieving uniform nanowire size, density, and alignment across the wafer remains a significant challenge, as small variations in growth parameters can lead to inconsistencies in the size and performance of each nanowire\cite{Church2022HolisticNanowires}. Overcoming these issues is vital for high output in the development of highly efficient energy conversion devices, including solar cells, photovoltaics, and photocatalysts\cite{Jia2019NanowireMacroscale}.

Recently, wafer-scale production of uniform GaAs/AlGaAs core-shell nanowires on two-inch Si(111) substrates using Ga-induced self-catalyzed MBE has been demonstrated\cite{Minehisa2023Wafer-scaleEpitaxy}. Assessing the performance and efficiency of these nanowires is essential, as achieving high yield across the entire wafer ensures consistent behavior and repeatability for light-emitting and light-detecting devices. By conducting close-to-single-nanowire measurements in micron-scale maps, we gain deeper insight into how local variations affect overall performance. In this work, we correlate growth parameters, such as nanowire density and vertical yield, observed through Scanning Electron Microscopy (SEM), with optoelectronic properties like photoluminescence (PL) intensity and carrier recombination lifetime, measured using high-throughput Time-Correlated Single-Photon Counting (TCSPC)\cite{Church2022HolisticNanowires}. Generating wafer and micron-scale maps of the two-inch substrate allowed us to observe inhomogeneities in PL emission at a fine scale while establishing uniform carrier lifetime across the wafer. Additionally, we assessed the absorptivity of the nanowires to illustrate their consistent performance and suitability for efficient solar energy applications, where high absorption is crucial. Such wafer-scale uniformity is particularly advantageous for commercial solar energy harvesting, promising enhanced performance and scalability\cite{Peng2011SiliconConversion}.


The GaAs core of these GaAs/AlGaAs core-shell nanowires was synthesized on a two-inch n-doped Si(111) substrate [Figure \ref{nanowire density}(a)] using the Ga-induced self-catalyzed Molecular Beam Epitaxy (MBE) technique, similar to the procedure described by \citet{Minehisa2023Wafer-scaleEpitaxy}. The nanowires consist of a nominally undoped GaAs core with a diameter of 100 nm, surrounded by a 5 nm thick intrinsic AlGaAs shell and a 75 nm thick p-doped AlGaAs shell containing 20\% Al and a doping density of 7$\times10^{18}$ cm$^{-3}$ [Figure \ref{nanowire density}(b)]. Furthermore, a 5 nm thick outer layer of p$^{+}$ doped GaAs with a doping density of 2$\times 10^{20}$ cm$^{-3}$ is grown. This outer layer of p$^{+}$ doped GaAs prevents oxidation of the AlGaAs surface, which can degrade performance by creating surface defects that act as nonrecombination centers\cite{Amato2016SurfaceNanowires}. Based on SEM measurements, the statistical mean diameter of 100 randomly selected nanowires is $255 \pm 24$ nm, and while the length is $5.30 \pm 0.36$ $\mu$m [Supplementary Figure S1]. The photograph in Figure \ref{nanowire density}(a) shows the consistent matte-black appearance of the wafer, which qualitatively demonstrates both the high absorptivity of the nanowires, and the uniform coverage across the wafer.

To quantify the structural uniformity of the growth, SEM observations at 130 points [Supplementary Figure S2,S3] across the 2-inch wafer sample were performed and processed to obtain the nanowire density and to measure the percentage of nanowires which are vertically orientated [Figure \ref{nanowire density}(c-e)] \cite{Fonseka2019EngineeringHeterostructures}. At the center of the wafer, the local nanowire density is $4.7 \times 10^8$ cm$^{-2}$, with a standard deviation of only 10\% across a diameter of 20 mm and 20\% across the entire 50 mm wafer. At the edge of the wafer, the density reduces to $3.5 \times 10^8$ cm$^{-2}$. The density is also skewed to one side of the wafer, which can be attributed to a slight inclination of the substrate holder, resulting in a temperature gradient on the surface. The vertical yield, defined as the percentage of nanowires which are vertically aligned, is largely correlated with the density, with values of 93.5\% and 83.4\%, at the center and edge of the wafer respectively. The mean vertical yield is 90\%, and varies by only 4\% across a 20 mm diameter and 7\% across the full wafer. Our self-catalyzed MBE approach achieves nanowire densities and yields comparable to those reported in high-density nanowire growth studies\cite{Fedorov2021TailoringSubstrates}, with the added advantage of uniformity across a two-inch wafer.


High absorptivity is crucial for efficient solar energy harvesting. To quantify this parameter, the reflectivity of the nanowires was measured using a diffused 532 nm laser and a step size of 1 mm in Figure \ref{nanowire density}(f) and the results were normalised at the nanowire-free edge of the wafer to the Fresnel reflectivity of Si, i.e., 0.37 [Supplemetary Figure S6(a)] \cite{Aspnes1983DielectricEV}. Assuming that the total of absorption and reflection sums to unity, we generated an absorption map from these reflection data, showing a mean absorption of 98\% across the wafer. The variation in absorptivity across the wafer was quantified by expanding the analysis from the wafer center, revealing a gradual increase in absorptivity variations with larger local regions, reaching an absorptivity variation of up to 1.2\% across the entire wafer [Supplementary Figure S7(a)].

During this experiment, PL emission from the nanowires was also detected. This emission was collected by a Horiba iHR550 spectrometer in order to measure the spectrum, which is shown in Figure \ref{TCSPC map}(a), where the spectrum of a GaAs wafer is also shown, for reference. The normalised spectra are very similar - with comparable peak wavelengths and spectra widths. The nanowire emission is therefore likely to originate from the nominally-undoped GaAs core. 

Evaluation of the optoelectronic performance and uniformity of the wafer is typically done by measuring the emission intensity, however, this is a convolution of many factors, such as the carrier density, recombination and light-extraction, as well as the nanowire density, and is therefore inefficient for evaluating the intrinsic nanowire quality. To address this, we employed a high-throughput experimental approach\cite{Church2022HolisticNanowires} that uses TCSPC to measure both emission intensity and carrier lifetime maps and correlates the density and quality of the nanowires at both wafer and micron scales. This was achieved by exciting the nanowires using a 150 fs pulsed laser with a wavelength of 515 nm and a 1 MHz repetition rate, which was focused to a spot of diameter of 1.4 $\mu$m. The emission was coupled into a Silicon-based single-photon avalanche diode and TCSPC histograms were generated by a Picoquant HydraHarp 400 timing system, using an integration time of 1 s for each measurement. Depending on the local nanowire density, this experiment will excite between 4 and 9 nanowires, on average, in each acquisition.

An example TCSPC decay is shown in Figure \ref{TCSPC map}(b), where two components are observed. A bi-exponential function is fit to the data to extract the carrier recombination lifetimes. The fast decay component has a lifetime of 0.1 ns, which is at the time response of this system, approximately 0.1 ns. This may be due to non-radiative recombination and remains constant across the wafer [Supplementary Figure S6(e,f)]\cite{Pevere2018RapidNanocrystals}. For this discussion we focus on the slower lifetime, of 1.7 ns. Due to the AlGaAs and GaAs passivation shell layers, this is unlikely to be due to surface recombination. Instead, these dynamics may result from a combination of radiative excitonic recombination and non-radiative Schockley-Reed-Hall (SRH) recombination related to point defects in the GaAs\cite{Niemeyer2019MeasurementGaAs}.

The TCSPC measurement was repeated for multiple micron-scale maps of 20 × 20 $\mu$m$^2$ local areas, measured at intervals of 1.75 mm across the full 50 mm wafer. From these local maps, a wafer-scale map was generated using the median and standard deviation values. The sensitivity of the technique was verified on a two-inch, 500 $\mu$m$^2$ thick single crystal (100) undoped GaAs wafer, which showed variations at least an order of magnitude smaller than those observed in the nanowire samples, confirming the method’s capability to detect subtle differences [Supplementary Figure S6(c,d)]. 

The PL intensity map and the carrier recombination lifetime maps are shown in Figure \ref{TCSPC map}(c,e), where each pixel value represents the median of the corresponding local maps, with example local maps shown in Figure \ref{TCSPC map}(d,f). Figure \ref{Variations}(a) shows the wafer-scale maps of local variation - calculated from the standard deviation of each micron-scale map, normalised to the median values.

The PL intensity is of comparable uniformity to the nanowire density on the wafer-scale - varying by only 10\% across the central 20 mm diameter, however, the variation increases to 35\% when considering the wafer as whole. This variation is skewed to the same side of the wafer as the density measurements in Figure \ref{nanowire density}(d), which is a strong indication of the impact that nanowire density has on the emission intensity. The micron-scale variation is lower on average, with a median value of 21\%, increasing slightly to 26 \% at the wafer edge: this suggests that the local uniformity does not change significantly across the wafer.

In contrast, the lifetime map in Figure \ref{TCSPC map}(e) is significantly more uniform, varying by only 2\% across a 40 mm diameter, increasing to 9 \% for the full wafer. This map does not show any evident skew corresponding to the nanowire density, indicating that carrier lifetime is less sensitive to density variations. Furthermore, the carrier lifetime variations at both the local [Figure \ref{Variations}(b)] and the wafer scales are similar, with median variations of 9\% in both cases, suggesting that the same process controls the uniformity in both regimes. This may be driven by changes in the non-radiative recombination rate, a result of variation in the density of defects between different the nanowires, the incorporation of which is related to different local growth conditions\cite{Furthmeier2014LongNanowires}. Alternatively, the variations could be excitation-related, such as the saturation of defect states with increased power\cite{Joyce2008HighCharacterization}. However, power-dependence measurements [Supplementary Figure S7(b)] were conducted prior to the experiments, ensuring that the measurements were performed in the linear regime, minimizing the impact of excitation saturation.

Further insight into the optoelectronic properties can be gained by exploring the relationships between the measured parameters by correlating these with the nanowire characteristics. This analysis reveals a relationship between nanowire density and vertical yield, PL intensity, and carrier lifetime, fitted to a power law model $(y=Ax^B)$ as shown in Figure \ref{Variations}(c-e). Figure \ref{Variations}(c) demonstrates that the regions of the wafer with a greater number of nanowires per unit area also have a greater vertical yield. Doubling the nanowire density increases the yield by 15\%, but this effect becomes less pronounced as the yield approaches 100\%, resulting in a sub-linear relationship. This observation has previously been explained by considering the surface contact between the nanowires and the substrate - which decreases at higher densities, and leads to a higher vertical yield, especially near the center of the wafer\cite{Plissard2011HighDropletpositioning}. 

Figure \ref{Variations}(d) shows that a higher nanowire density also leads to a higher emission intensity. This can be understood as a higher density leads to more nanowires, on average, being within the excitation spot and therefore more GaAs material emitting during each acquisition\cite{KimInfluenceNanowires}. In this explanation, the emission intensity would increase linearly with nanowire density, however, Figure \ref{Variations}(d) demonstrates that this relationship is sublinear. Fitting the intensity data with the power law model $(y=Ax^B)$ yields a power factor of $B = 0.5$, shown by the red line in Figure \ref{Variations}(d). This result implies that the overall emission efficiency is dropping at higher nanowire densities, and drops by 12\% when compared with low density. 

The TCSPC allows changes in the Internal Quantum Efficiency (IQE) to be correlated with these effects. This metric can be accessed through the carrier lifetime, which is shown in Figure \ref{Variations}(e). Here, regions with higher nanowire densities tend to have longer carrier lifetimes - which is likely caused by a reduced non-radiative recombination rate. This suggests that the local growth conditions which nucleate more nanowires also lead to a local reduction in the defect density within the nanowires. However, this effect is small - with only a 6\% change in the lifetime across the density range, and so it is unlikely to be the cause of the intensity variation.

The PL intensity may instead be linked with the vertical yield, since light coupling varies upon the orientation of the nanowires, and this will impact the overall coverage of the substrate. Nanowires which are vertically aligned will expose the minimum surface area to the excitation and, in the most limiting case of a horizontal nanowire, the exposed surface area will increase by a factor of 6 (using the nominal nanowire width, 0.3 $\mu$m, and the excitation spot diameter, 1.4 $\mu$m). In this simple picture [Figure \ref{Variations}(f)], an increase in the vertical yield from 80\% to 95\% reduces the substrate coverage by 35\%. As the absorption depth in GaAs at 532 nm is approximately 200–300 nm\cite{Casey1975ConcentrationEV}, smaller than the dimensions of the nanowire, the volume of material probed by the excitation will scale linearly with this area. Therefore, assuming that the light coupling efficiency is similar from each facet, a drop in the apparent efficiency by a similar amount would be expected. 

In practice, non-vertical nanowires are at an angle with respect to the substrate, and substrate coverage will be smaller than this calculation. A reasonable estimate can be found from the SEM results, such as those in Supplementary Figure S5, where an analysis of the object size in the image puts the median projection length at 0.4 $\mu$m, which is independent of the position on the wafer. This results in a 10\% reduction in the effective area of the substrate covered by the nanowires, indicating a decrease in the substrate’s effective absorption fraction across the wafer. The magnitude of this effect is therefore compatible with the observations in Figure \ref{Variations}(d) and is likely the factor that drives the sub-linearity of the PL intensity with nanowire density.

In conclusion, this study highlights the successful synthesis and detailed characterization of GaAs/AlGaAs core-shell nanowires on a two-inch n-doped Si(111) substrate using Ga-induced self-catalyzed MBE. Comprehensive high-throughput characterization, including nanowire density and vertical yield maps from SEM and PL intensity and carrier recombination lifetime maps from TCSPC measurements, revealed a nanowire density of $4.7 \times 10^8$ cm$^{-2}$ at the wafer centre, and a mean vertical yield of $90\pm7\%$. The nanowires are grown with excellent uniformity, with these parameters varying by only 10\% and 4\% across the central 20 mm of the wafer. In this same area, the PL intensity and carrier lifetime vary by 10\% and 2\%. The lifetime showed a consistent variation on both the wafer and micrometer scales, demonstrating that this measurement is decoupled from nanowire density and pumping efficiency effects, and resulting in a similar recombination efficiency throughout the wafer. In contrast, the PL intensity exhibits greater variability, with 35\% variation at the wafer scale and 21\% on the micron scale. This variation is tied to local and wafer-scale changes in nanowire density, along with variation in nanowire orientation - which impacts the effective pumping volume. In general, the regions of the wafer with higher nanowire densities also have higher vertical yields, PL intensities and recombination efficiencies. The overall uniformity and quality of the nanowires is promising for large-scale applications, particularly in optoelectronics and solar energy harvesting and these findings pave the way for scalable applications of III-V semiconductor nanowires on inexpensive silicon substrates in advanced optoelectronic devices, such as high-efficiency solar cells and photodetectors, where both material quality and uniformity are paramount.

\begin{acknowledgments}
This work was funded by EPSRC under grant MR/T021519/1, the Leverhulme Trust under Fellowship ECF-2024-250 and by KAKENHI (Nos. 23H00250, 21KK0068, 19H00855) from the Japan Society of Promotion of Science

\end{acknowledgments}

\section*{References}

%

\renewcommand{\thefigure}{S\arabic{figure}}
\renewcommand{\thetable}{S\arabic{table}}
\renewcommand{\theequation}{S\arabic{equation}}
\setcounter{figure}{0} 
\setcounter{equation}{0} 
\setcounter{table}{0}

\clearpage  

\begin{figure*}[h]
    \centering
    \includegraphics[width=1\linewidth]{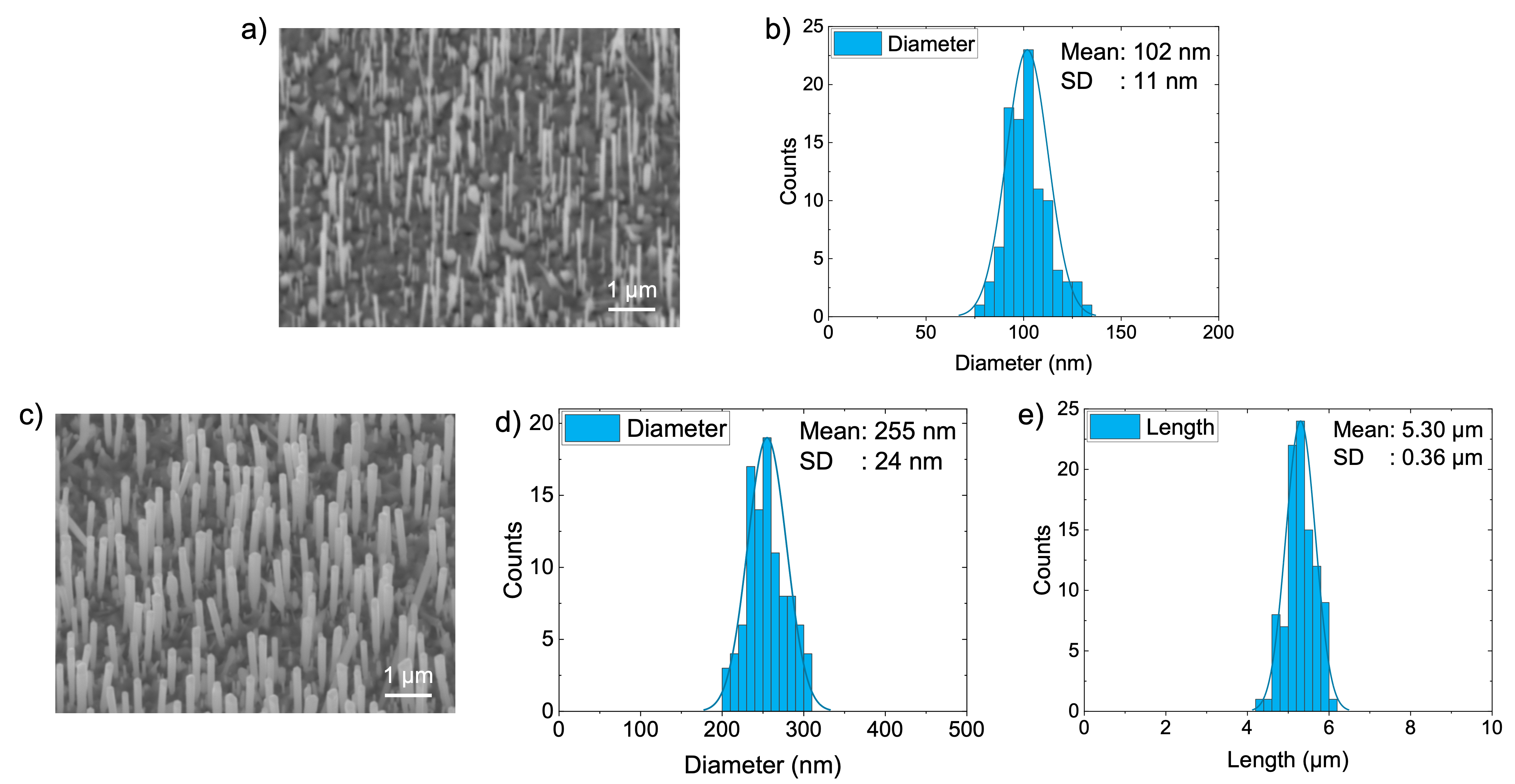}
    \caption{(a) 20$^{\circ}$ tilted SEM image of GaAs core nanowires, used to determine core diameter. (b) Histogram showing the statistical mean diameter of 100 randomly selected GaAs nanowires from (a). (c) 20$^{\circ}$ tilted SEM image of GaAs/AlGaAs core-shell nanowires. (d) Histogram showing the statistical mean core-shell diameter, and (e) histogram showing the statistical mean length of 100 randomly selected core-shell nanowires from (c).}
    \label{dimensions}
\end{figure*}

\begin{figure*}[h]
    \centering
    \includegraphics[width=1\linewidth]{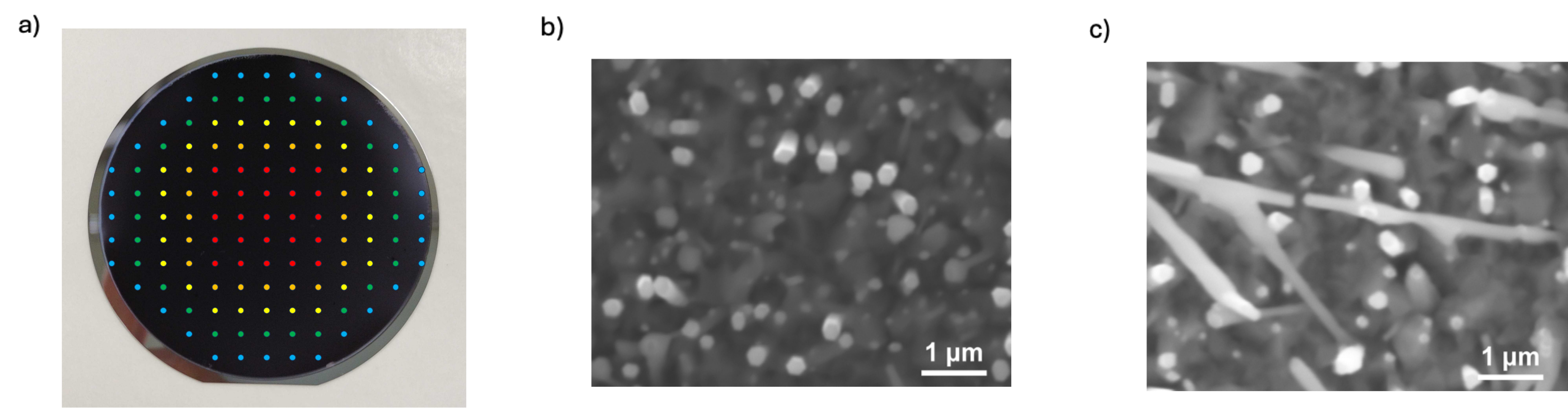}
    \caption{(a) A two-inch wafer containing GaAs/AlGaAs core-shell nanowires with 130 points marked for estimating nanowire density and vertical yield. The marked areas include the center (red circle), edge 1 (orange circle), edge 2 (yellow circle), edge 3 (green circle), and edge 4 or the outermost area (blue circle). (b) SEM image of the wafer’s center, showing high nanowire density and vertical yield. (c) SEM image of the wafer’s edge, showing lower nanowire density and vertical yield.}
    \label{SEM}
\end{figure*}

\begin{figure*}[h]
    \centering
    \includegraphics[width=1\linewidth]{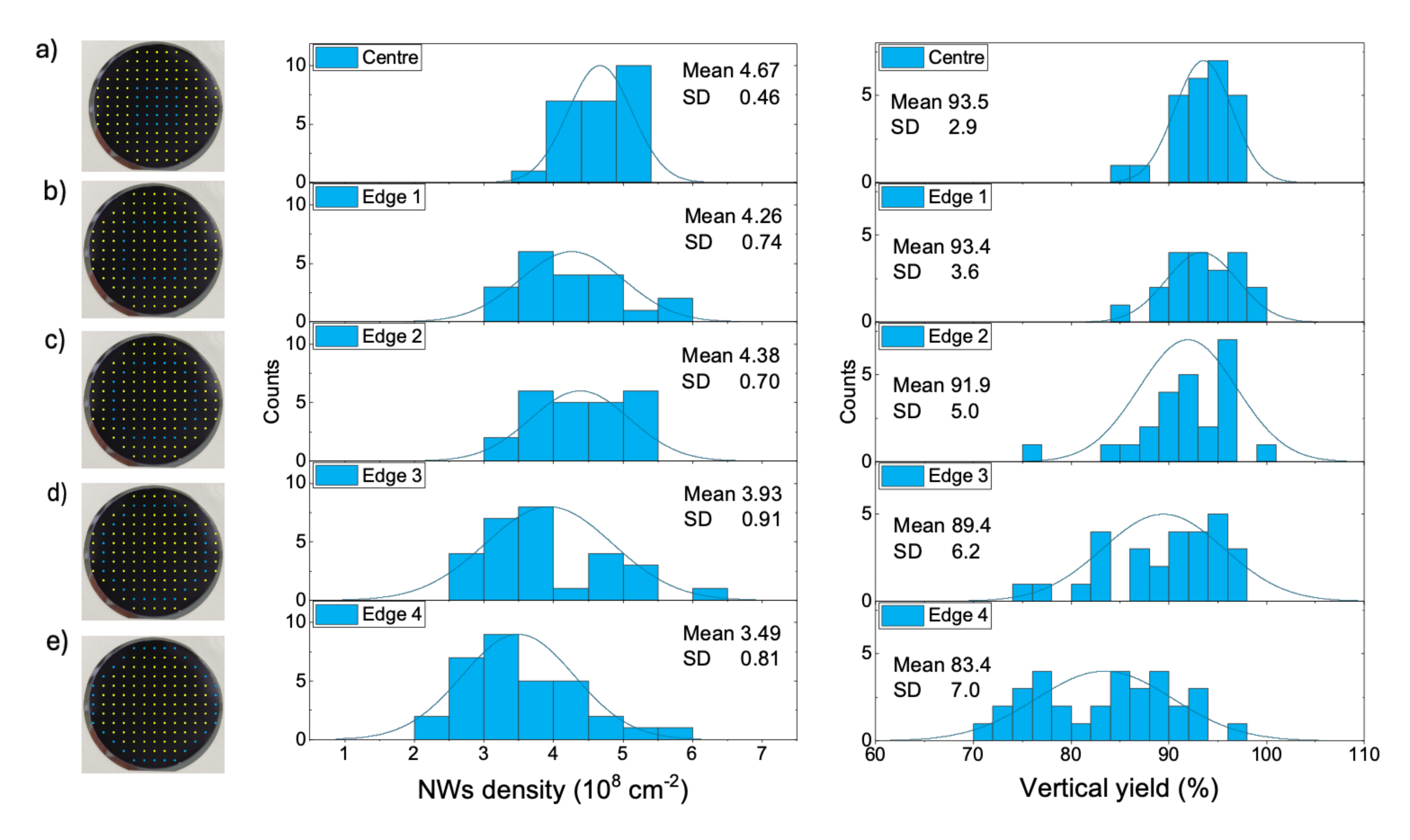}
    \caption{Area specific statictics. Density at centre: $(4.67 \pm 0.85) \times 10^8$ cm$^{-2}$, Density at edge: $(3.49 \pm 0.85) \times 10^8$ cm$^{-2}$. Vertical Yield at centre: 93.5\%, Vertical yield at edge: 83.4\%}
    \label{SEM stats}
\end{figure*}

\begin{figure*}[h]
    \centering
    \includegraphics[width=1\linewidth]{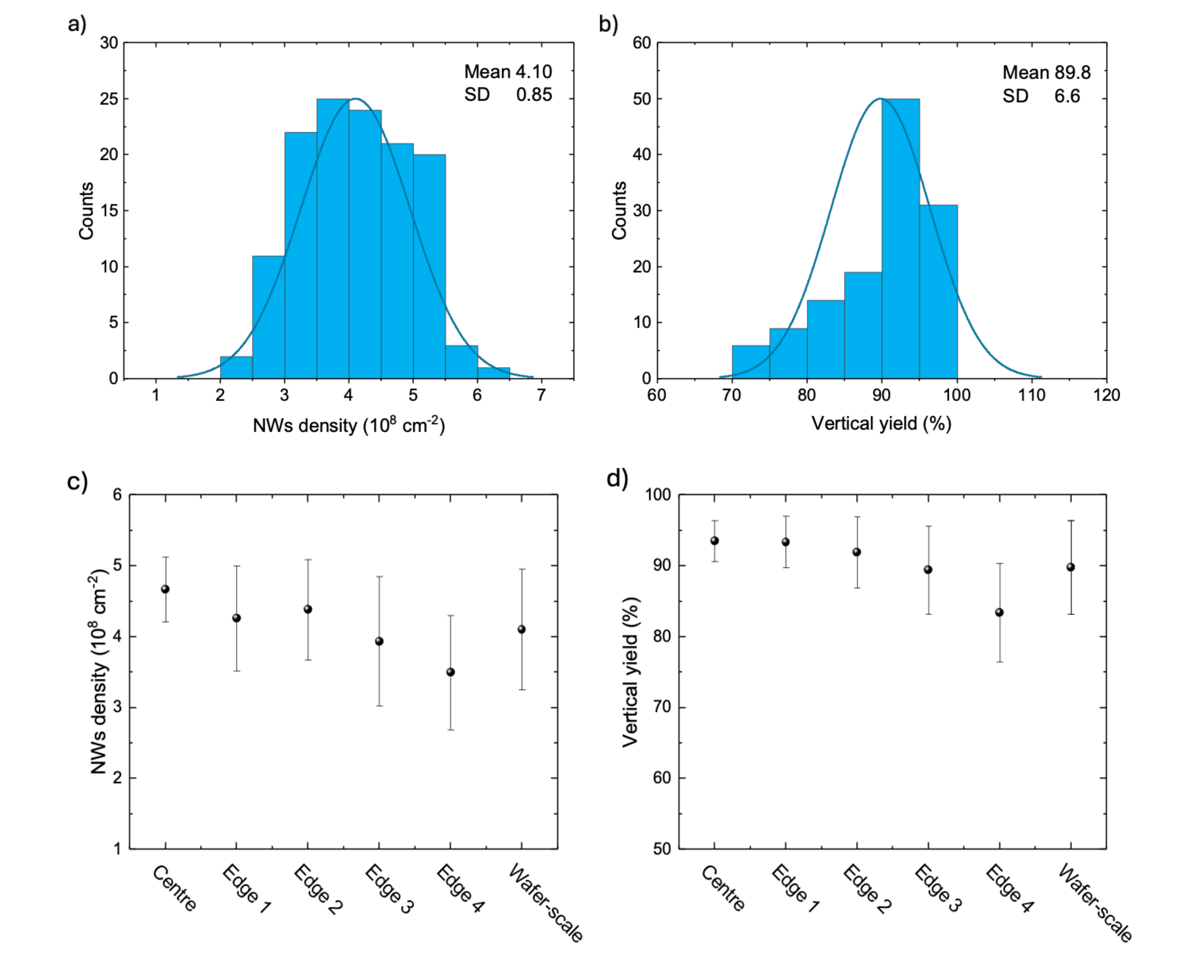}
    \caption{The figures above present wafer-scale statistics. (a) shows the mean nanowire density of $4.1 \times 10^8$ cm$^{-2}$ with a standard deviation of $0.85 \times 10^8$ cm$^{-2}$. (b) displays the mean vertical yield of 89.8\% with a standard deviation of 6.6\%. (c) and (d) illustrate the decreasing trend in nanowire density and vertical yield from the center to the edge of the wafer. Both nanowire density and vertical yield exhibit greater standard deviation as the distance from the center to the edge increases.}
    \label{wafer stats}
\end{figure*}

\begin{figure*}[h]
    \centering
    \includegraphics[width=1\linewidth]{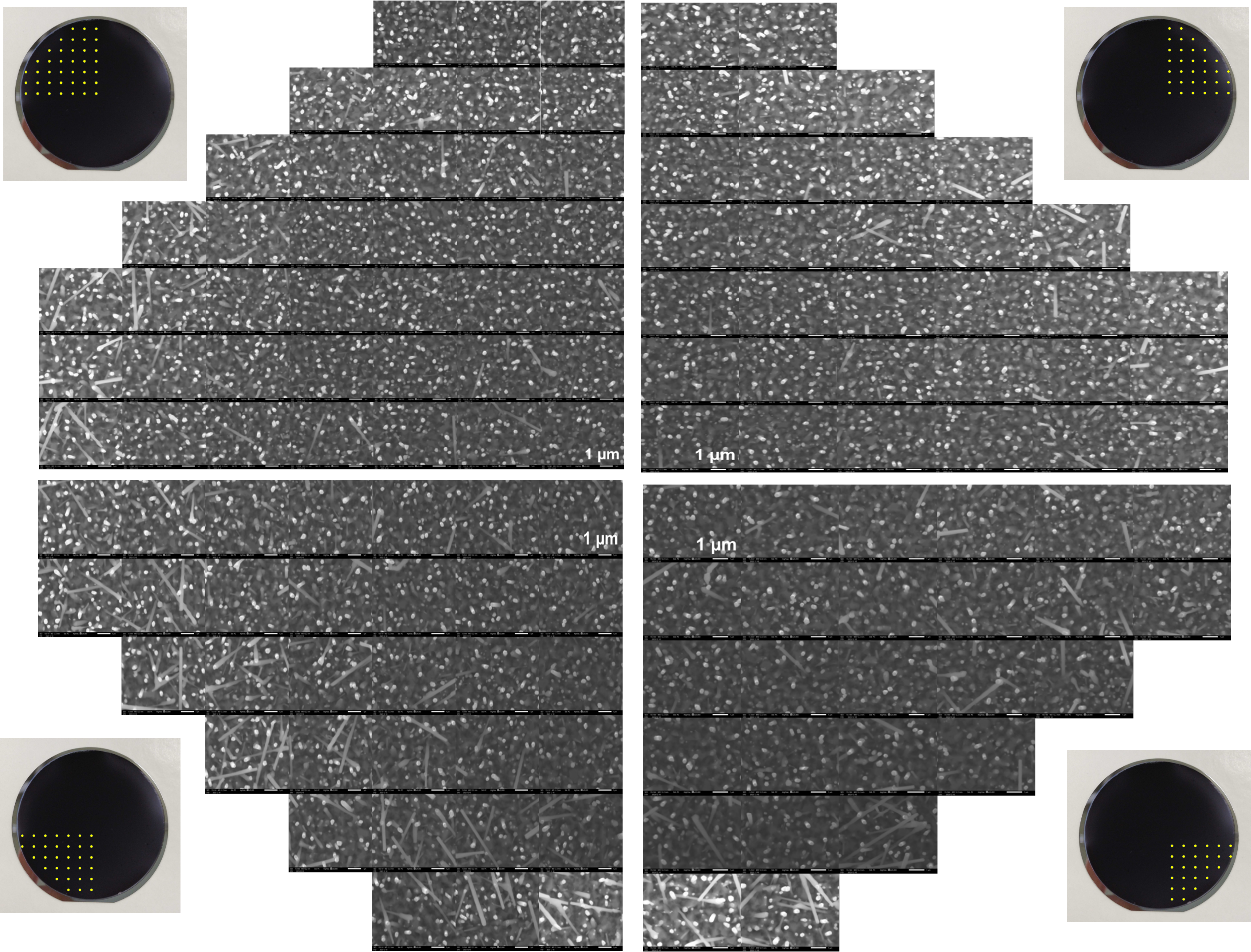}
    \caption{Composite SEM images of the entire two-inch wafer, showing the gradual transition from high nanowire density and vertical yield at the center to lower density and yield toward the edges.}
    \label{SEM image}
\end{figure*} 

\begin{figure*}[h]
    \centering
    \includegraphics[width=1\linewidth]{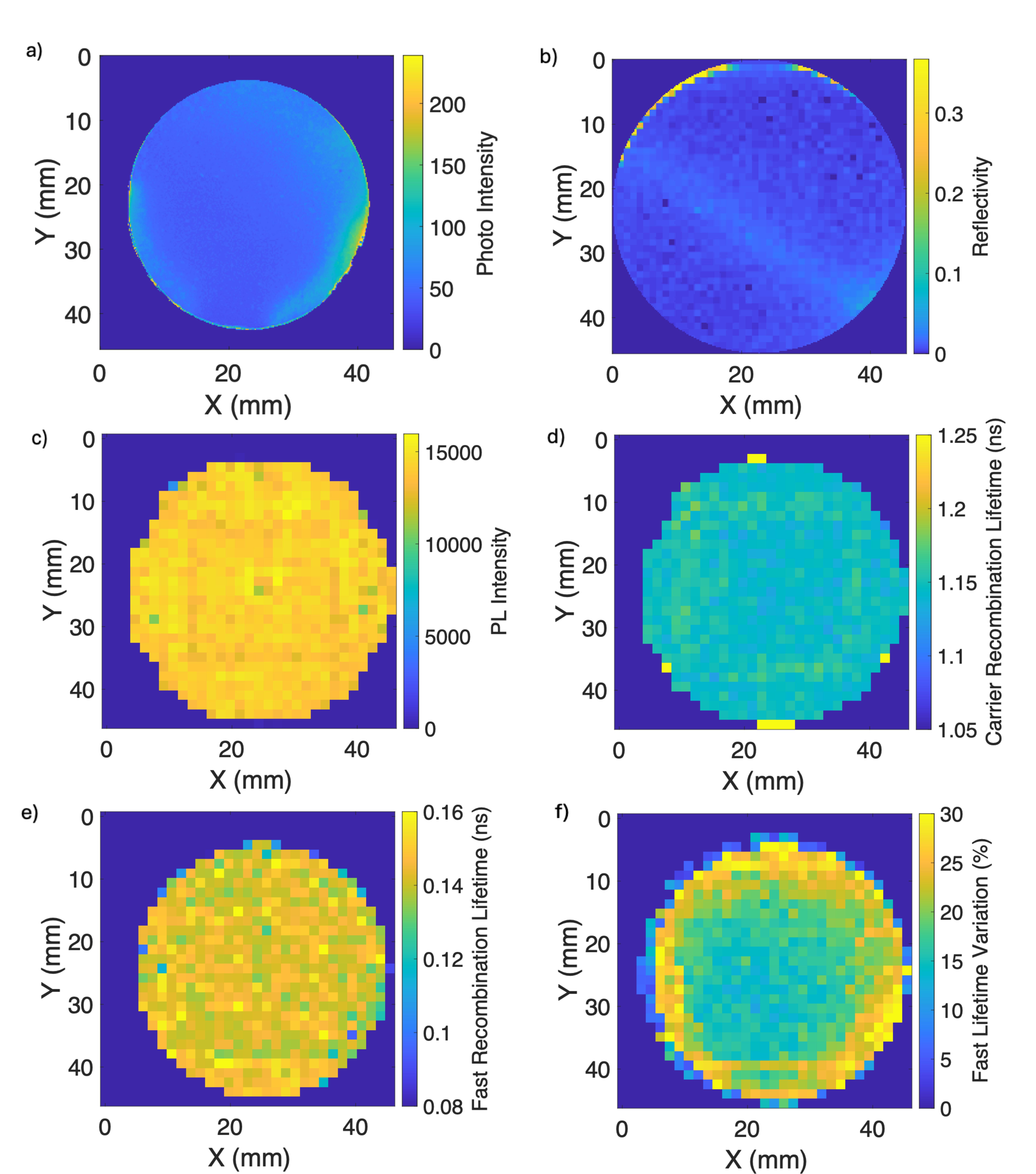}
    \caption{(a) Photo Intensity Map, generated by performing image registration of \ref{SEM}(a) over TCSPC and SEM maps, highlighting maximum absorption (pitch black) and discolouration in specific regions. (b) Reflectivity Map produced by normalizing the maximum reflection region to the Fresnel reflection of silicon, where the edge of the wafer contains silicon. (c) and (d) Photoluminescence Intensity Map and Carrier Recombination Lifetime Map of a 500 $\mu$m$^2$ thick single-crystal (100) undoped GaAs reference sample, respectively, used to verify the experimental technique. (e) Fast Recombination Lifetime Map of the GaAs nanowire wafer, and (f) its Local Variation Map.}
    \label{More maps!}
\end{figure*}

\begin{figure*}[h]
    \centering
    \includegraphics[width=1\linewidth]{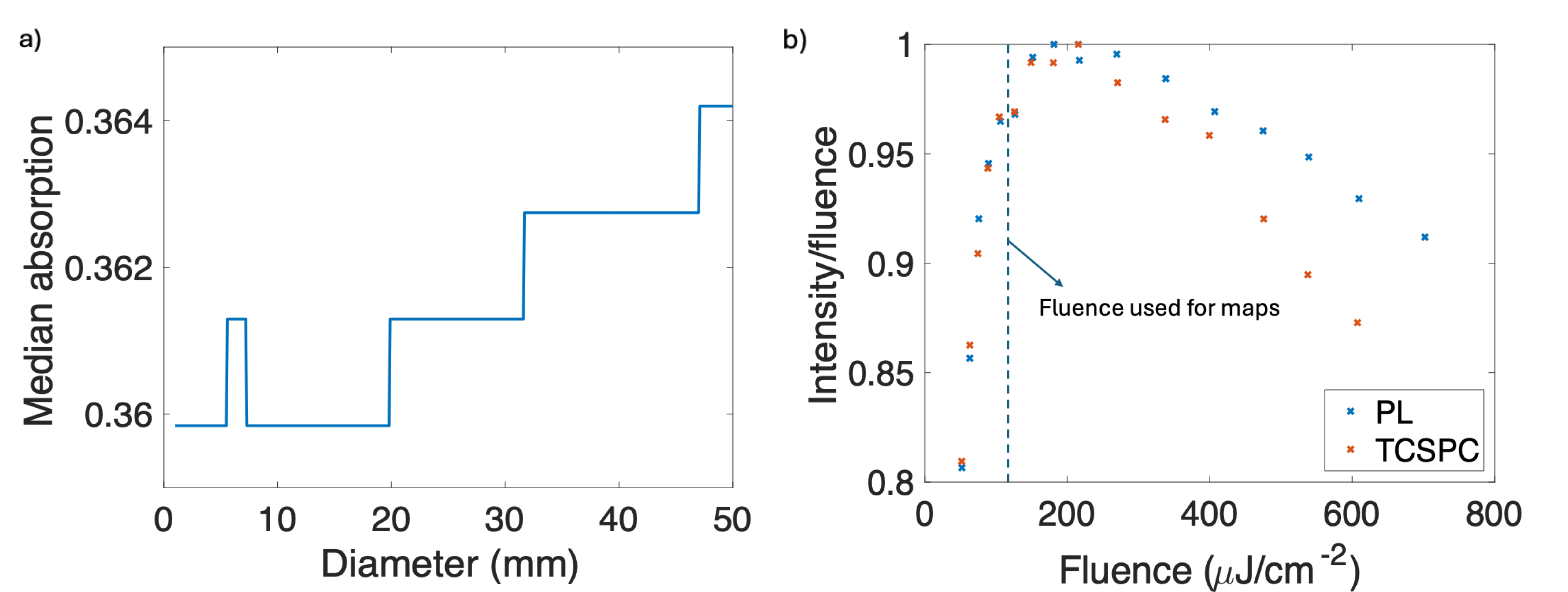}
mi    \caption{(a) Median absorptivity as a function of wafer diameter, measured radially from the center, demonstrating uniformity across the wafer with slight variations in absorptivity. (b) Power-dependence measurements showing intensity normalized by fluence  against fluence, confirming that the experiments were conducted in the linear regime, thereby minimizing the impact of excitation saturation.}
    \label{More maps!}
\end{figure*}

\begin{figure*}[h]
    \centering
    \includegraphics[width=1\linewidth]{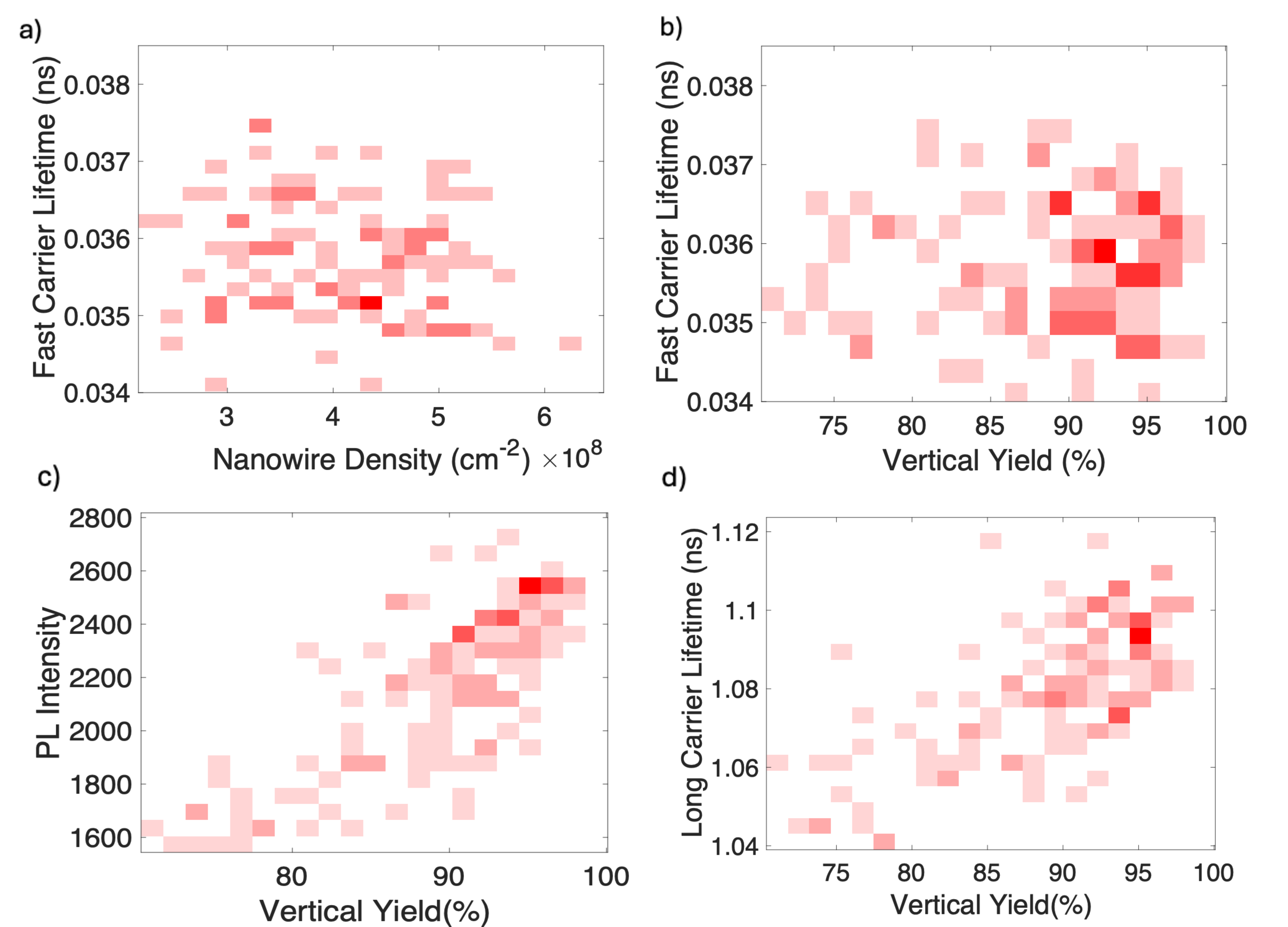}
    \caption{Correlation plots showing: (a) Nanowire density vs. fast carrier lifetime, indicating weak correlation; (b) Vertical yield vs. fast carrier lifetime, indicating weak correlation; (c) Vertical yield vs. PL intensity, showing higher intensity with increased vertical yield; (d) Vertical yield vs. carrier recombination lifetime, also showing higher lifetime with increased vertical yield.}
    \label{correlations}
\end{figure*} 


\begin{thebibliography}{31}%
\makeatletter
\providecommand \@ifxundefined [1]{%
 \@ifx{#1\undefined}
}%
\providecommand \@ifnum [1]{%
 \ifnum #1\expandafter \@firstoftwo
 \else \expandafter \@secondoftwo
 \fi
}%
\providecommand \@ifx [1]{%
 \ifx #1\expandafter \@firstoftwo
 \else \expandafter \@secondoftwo
 \fi
}%
\providecommand \natexlab [1]{#1}%
\providecommand \enquote  [1]{``#1''}%
\providecommand \bibnamefont  [1]{#1}%
\providecommand \bibfnamefont [1]{#1}%
\providecommand \citenamefont [1]{#1}%
\providecommand \href@noop [0]{\@secondoftwo}%
\providecommand \href [0]{\begingroup \@sanitize@url \@href}%
\providecommand \@href[1]{\@@startlink{#1}\@@href}%
\providecommand \@@href[1]{\endgroup#1\@@endlink}%
\providecommand \@sanitize@url [0]{\catcode `\\12\catcode `\$12\catcode `\&12\catcode `\#12\catcode `\^12\catcode `\_12\catcode `\%12\relax}%
\providecommand \@@startlink[1]{}%
\providecommand \@@endlink[0]{}%
\providecommand \url  [0]{\begingroup\@sanitize@url \@url }%
\providecommand \@url [1]{\endgroup\@href {#1}{\urlprefix }}%
\providecommand \urlprefix  [0]{URL }%
\providecommand \Eprint [0]{\href }%
\providecommand \doibase [0]{http://dx.doi.org/}%
\providecommand \selectlanguage [0]{\@gobble}%
\providecommand \bibinfo  [0]{\@secondoftwo}%
\providecommand \bibfield  [0]{\@secondoftwo}%
\providecommand \translation [1]{[#1]}%
\providecommand \BibitemOpen [0]{}%
\providecommand \bibitemStop [0]{}%
\providecommand \bibitemNoStop [0]{.\EOS\space}%
\providecommand \EOS [0]{\spacefactor3000\relax}%
\providecommand \BibitemShut  [1]{\csname bibitem#1\endcsname}%
\let\auto@bib@innerbib\@empty
\bibitem [{\citenamefont {M{\aa}rtensson}\ \emph {et~al.}(2004)\citenamefont {M{\aa}rtensson}, \citenamefont {Svensson}, \citenamefont {Wacaser}, \citenamefont {Larsson}, \citenamefont {Seifert}, \citenamefont {Deppert}, \citenamefont {Gustafsson}, \citenamefont {Wallenberg},\ and\ \citenamefont {Samuelson}}]{Martensson2004EpitaxialSilicon}%
  \BibitemOpen
  \bibfield  {author} {\bibinfo {author} {\bibfnamefont {T.}~\bibnamefont {M{\aa}rtensson}}, \bibinfo {author} {\bibfnamefont {C.~P.~T.}\ \bibnamefont {Svensson}}, \bibinfo {author} {\bibfnamefont {B.~A.}\ \bibnamefont {Wacaser}}, \bibinfo {author} {\bibfnamefont {M.~W.}\ \bibnamefont {Larsson}}, \bibinfo {author} {\bibfnamefont {W.}~\bibnamefont {Seifert}}, \bibinfo {author} {\bibfnamefont {K.}~\bibnamefont {Deppert}}, \bibinfo {author} {\bibfnamefont {A.}~\bibnamefont {Gustafsson}}, \bibinfo {author} {\bibfnamefont {L.~R.}\ \bibnamefont {Wallenberg}}, \ and\ \bibinfo {author} {\bibfnamefont {L.}~\bibnamefont {Samuelson}},\ }\bibfield  {title} {\enquote {\bibinfo {title} {{Epitaxial III-V nanowires on silicon}},}\ }\href {\doibase 10.1021/NL0487267} {\bibfield  {journal} {\bibinfo  {journal} {Nano Letters}\ }\textbf {\bibinfo {volume} {4}},\ \bibinfo {pages} {1987--1990} (\bibinfo {year} {2004})}\BibitemShut {NoStop}%
\bibitem [{\citenamefont {Chuang}\ \emph {et~al.}(2007)\citenamefont {Chuang}, \citenamefont {Moewe}, \citenamefont {Chase}, \citenamefont {Kobayashi}, \citenamefont {Chang-Hasnain},\ and\ \citenamefont {Crankshaw}}]{Chuang2007CriticalSubstrates}%
  \BibitemOpen
  \bibfield  {author} {\bibinfo {author} {\bibfnamefont {L.~C.}\ \bibnamefont {Chuang}}, \bibinfo {author} {\bibfnamefont {M.}~\bibnamefont {Moewe}}, \bibinfo {author} {\bibfnamefont {C.}~\bibnamefont {Chase}}, \bibinfo {author} {\bibfnamefont {N.~P.}\ \bibnamefont {Kobayashi}}, \bibinfo {author} {\bibfnamefont {C.}~\bibnamefont {Chang-Hasnain}}, \ and\ \bibinfo {author} {\bibfnamefont {S.}~\bibnamefont {Crankshaw}},\ }\bibfield  {title} {\enquote {\bibinfo {title} {{Critical diameter for III-V nanowires grown on lattice-mismatched substrates}},}\ }\href {\doibase 10.1063/1.2436655/333529} {\bibfield  {journal} {\bibinfo  {journal} {Applied Physics Letters}\ }\textbf {\bibinfo {volume} {90}},\ \bibinfo {pages} {43115} (\bibinfo {year} {2007})}\BibitemShut {NoStop}%
\bibitem [{\citenamefont {Borg}\ \emph {et~al.}(2014)\citenamefont {Borg}, \citenamefont {Schmid}, \citenamefont {Moselund}, \citenamefont {Signorello}, \citenamefont {Gignac}, \citenamefont {Bruley}, \citenamefont {Breslin}, \citenamefont {Das~Kanungo}, \citenamefont {Werner},\ and\ \citenamefont {Riel}}]{Borg2014VerticalSi100}%
  \BibitemOpen
  \bibfield  {author} {\bibinfo {author} {\bibfnamefont {M.}~\bibnamefont {Borg}}, \bibinfo {author} {\bibfnamefont {H.}~\bibnamefont {Schmid}}, \bibinfo {author} {\bibfnamefont {K.~E.}\ \bibnamefont {Moselund}}, \bibinfo {author} {\bibfnamefont {G.}~\bibnamefont {Signorello}}, \bibinfo {author} {\bibfnamefont {L.}~\bibnamefont {Gignac}}, \bibinfo {author} {\bibfnamefont {J.}~\bibnamefont {Bruley}}, \bibinfo {author} {\bibfnamefont {C.}~\bibnamefont {Breslin}}, \bibinfo {author} {\bibfnamefont {P.}~\bibnamefont {Das~Kanungo}}, \bibinfo {author} {\bibfnamefont {P.}~\bibnamefont {Werner}}, \ and\ \bibinfo {author} {\bibfnamefont {H.}~\bibnamefont {Riel}},\ }\bibfield  {title} {\enquote {\bibinfo {title} {{Vertical III-V nanowire device integration on Si(100)}},}\ }\href {\doibase 10.1021/NL404743J/SUPPL{\_}FILE/NL404743J{\_}SI{\_}001.PDF} {\bibfield  {journal} {\bibinfo  {journal} {Nano Letters}\ }\textbf {\bibinfo {volume} {14}},\ \bibinfo {pages} {1914--1920} (\bibinfo {year} {2014})}\BibitemShut {NoStop}%
\bibitem [{\citenamefont {Goktas}\ \emph {et~al.}(2018)\citenamefont {Goktas}, \citenamefont {Wilson}, \citenamefont {Ghukasyan}, \citenamefont {Wagner}, \citenamefont {McNamee},\ and\ \citenamefont {LaPierre}}]{Goktas2018NanowiresReview}%
  \BibitemOpen
  \bibfield  {author} {\bibinfo {author} {\bibfnamefont {N.~I.}\ \bibnamefont {Goktas}}, \bibinfo {author} {\bibfnamefont {P.}~\bibnamefont {Wilson}}, \bibinfo {author} {\bibfnamefont {A.}~\bibnamefont {Ghukasyan}}, \bibinfo {author} {\bibfnamefont {D.}~\bibnamefont {Wagner}}, \bibinfo {author} {\bibfnamefont {S.}~\bibnamefont {McNamee}}, \ and\ \bibinfo {author} {\bibfnamefont {R.~R.}\ \bibnamefont {LaPierre}},\ }\bibfield  {title} {\enquote {\bibinfo {title} {{Nanowires for energy: A review}},}\ }\href {\doibase 10.1063/1.5054842/124236} {\bibfield  {journal} {\bibinfo  {journal} {Applied Physics Reviews}\ }\textbf {\bibinfo {volume} {5}},\ \bibinfo {pages} {41305} (\bibinfo {year} {2018})}\BibitemShut {NoStop}%
\bibitem [{\citenamefont {Kelzenberg}\ \emph {et~al.}(2010)\citenamefont {Kelzenberg}, \citenamefont {Boettcher}, \citenamefont {Petykiewicz}, \citenamefont {Turner-Evans}, \citenamefont {Putnam}, \citenamefont {Warren}, \citenamefont {Spurgeon}, \citenamefont {Briggs}, \citenamefont {Lewis},\ and\ \citenamefont {Atwater}}]{Kelzenberg2010EnhancedApplications}%
  \BibitemOpen
  \bibfield  {author} {\bibinfo {author} {\bibfnamefont {M.~D.}\ \bibnamefont {Kelzenberg}}, \bibinfo {author} {\bibfnamefont {S.~W.}\ \bibnamefont {Boettcher}}, \bibinfo {author} {\bibfnamefont {J.~A.}\ \bibnamefont {Petykiewicz}}, \bibinfo {author} {\bibfnamefont {D.~B.}\ \bibnamefont {Turner-Evans}}, \bibinfo {author} {\bibfnamefont {M.~C.}\ \bibnamefont {Putnam}}, \bibinfo {author} {\bibfnamefont {E.~L.}\ \bibnamefont {Warren}}, \bibinfo {author} {\bibfnamefont {J.~M.}\ \bibnamefont {Spurgeon}}, \bibinfo {author} {\bibfnamefont {R.~M.}\ \bibnamefont {Briggs}}, \bibinfo {author} {\bibfnamefont {N.~S.}\ \bibnamefont {Lewis}}, \ and\ \bibinfo {author} {\bibfnamefont {H.~A.}\ \bibnamefont {Atwater}},\ }\bibfield  {title} {\enquote {\bibinfo {title} {{Enhanced absorption and carrier collection in Si wire arrays for photovoltaic applications}},}\ }\href {\doibase 10.1038/NMAT2635} {\ ,\ \bibinfo {pages} {14} (\bibinfo {year} {2010})}\BibitemShut {NoStop}%
\bibitem [{\citenamefont {Muskens}\ \emph {et~al.}(2008)\citenamefont {Muskens}, \citenamefont {Rivas}, \citenamefont {Algra}, \citenamefont {Bakkers},\ and\ \citenamefont {Lagendijk}}]{Muskens2008DesignApplications}%
  \BibitemOpen
  \bibfield  {author} {\bibinfo {author} {\bibfnamefont {O.~L.}\ \bibnamefont {Muskens}}, \bibinfo {author} {\bibfnamefont {J.~G.}\ \bibnamefont {Rivas}}, \bibinfo {author} {\bibfnamefont {R.~E.}\ \bibnamefont {Algra}}, \bibinfo {author} {\bibfnamefont {E.~P.}\ \bibnamefont {Bakkers}}, \ and\ \bibinfo {author} {\bibfnamefont {A.}~\bibnamefont {Lagendijk}},\ }\bibfield  {title} {\enquote {\bibinfo {title} {{Design of light scattering in nanowire materials for photovoltaic applications}},}\ }\href {\doibase 10.1021/NL0808076/ASSET/IMAGES/LARGE/NL-2008-008076{\_}0005.JPEG} {\bibfield  {journal} {\bibinfo  {journal} {Nano Letters}\ }\textbf {\bibinfo {volume} {8}},\ \bibinfo {pages} {2638--2642} (\bibinfo {year} {2008})}\BibitemShut {NoStop}%
\bibitem [{\citenamefont {Diedenhofen}\ \emph {et~al.}(2011)\citenamefont {Diedenhofen}, \citenamefont {Janssen}, \citenamefont {Grzela}, \citenamefont {Bakkers},\ and\ \citenamefont {G{\'{o}}mez~Rivas}}]{Diedenhofen2011StrongNanowires}%
  \BibitemOpen
  \bibfield  {author} {\bibinfo {author} {\bibfnamefont {S.~L.}\ \bibnamefont {Diedenhofen}}, \bibinfo {author} {\bibfnamefont {O.~T.}\ \bibnamefont {Janssen}}, \bibinfo {author} {\bibfnamefont {G.}~\bibnamefont {Grzela}}, \bibinfo {author} {\bibfnamefont {E.~P.}\ \bibnamefont {Bakkers}}, \ and\ \bibinfo {author} {\bibfnamefont {J.}~\bibnamefont {G{\'{o}}mez~Rivas}},\ }\bibfield  {title} {\enquote {\bibinfo {title} {{Strong geometrical dependence of the absorption of light in arrays of semiconductor nanowires}},}\ }\href {\doibase 10.1021/NN103596N/SUPPL{\_}FILE/NN103596N{\_}SI{\_}001.PDF} {\bibfield  {journal} {\bibinfo  {journal} {ACS Nano}\ }\textbf {\bibinfo {volume} {5}},\ \bibinfo {pages} {2316--2323} (\bibinfo {year} {2011})}\BibitemShut {NoStop}%
\bibitem [{\citenamefont {Mata}\ \emph {et~al.}(2013)\citenamefont {Mata}, \citenamefont {Zhou}, \citenamefont {Furtmayr}, \citenamefont {Teubert}, \citenamefont {Grade{\v{c}}ak}, \citenamefont {Eickhoff}, \citenamefont {Fontcuberta I~Morral},\ and\ \citenamefont {Arbiol}}]{Mata2013ANanowire}%
  \BibitemOpen
  \bibfield  {author} {\bibinfo {author} {\bibfnamefont {M.~D.~L.}\ \bibnamefont {Mata}}, \bibinfo {author} {\bibfnamefont {X.}~\bibnamefont {Zhou}}, \bibinfo {author} {\bibfnamefont {F.}~\bibnamefont {Furtmayr}}, \bibinfo {author} {\bibfnamefont {J.}~\bibnamefont {Teubert}}, \bibinfo {author} {\bibfnamefont {S.}~\bibnamefont {Grade{\v{c}}ak}}, \bibinfo {author} {\bibfnamefont {M.}~\bibnamefont {Eickhoff}}, \bibinfo {author} {\bibfnamefont {A.}~\bibnamefont {Fontcuberta I~Morral}}, \ and\ \bibinfo {author} {\bibfnamefont {J.}~\bibnamefont {Arbiol}},\ }\bibfield  {title} {\enquote {\bibinfo {title} {{A review of MBE grown 0D, 1D and 2D quantum structures in a nanowire}},}\ }\href {\doibase 10.1039/C3TC30556B} {\bibfield  {journal} {\bibinfo  {journal} {Journal of Materials Chemistry C}\ }\textbf {\bibinfo {volume} {1}},\ \bibinfo {pages} {4300--4312} (\bibinfo {year} {2013})}\BibitemShut {NoStop}%
\bibitem [{\citenamefont {Gudiksen}\ \emph {et~al.}(2002)\citenamefont {Gudiksen}, \citenamefont {Lauhon}, \citenamefont {Wang}, \citenamefont {Smith},\ and\ \citenamefont {Lieber}}]{Gudiksen2002GrowthElectronics}%
  \BibitemOpen
  \bibfield  {author} {\bibinfo {author} {\bibfnamefont {M.~S.}\ \bibnamefont {Gudiksen}}, \bibinfo {author} {\bibfnamefont {L.~J.}\ \bibnamefont {Lauhon}}, \bibinfo {author} {\bibfnamefont {J.}~\bibnamefont {Wang}}, \bibinfo {author} {\bibfnamefont {D.~C.}\ \bibnamefont {Smith}}, \ and\ \bibinfo {author} {\bibfnamefont {C.~M.}\ \bibnamefont {Lieber}},\ }\bibfield  {title} {\enquote {\bibinfo {title} {{Growth of nanowire superlattice structures for nanoscale photonics and electronics}},}\ }\href {\doibase 10.1038/415617a} {\bibfield  {journal} {\bibinfo  {journal} {Nature 2002 415:6872}\ }\textbf {\bibinfo {volume} {415}},\ \bibinfo {pages} {617--620} (\bibinfo {year} {2002})}\BibitemShut {NoStop}%
\bibitem [{\citenamefont {Chen}\ \emph {et~al.}(2015)\citenamefont {Chen}, \citenamefont {Chen}, \citenamefont {Ishikawa},\ and\ \citenamefont {Buyanova}}]{Chen2015SuppressionNanowires}%
  \BibitemOpen
  \bibfield  {author} {\bibinfo {author} {\bibfnamefont {S.~L.}\ \bibnamefont {Chen}}, \bibinfo {author} {\bibfnamefont {W.~M.}\ \bibnamefont {Chen}}, \bibinfo {author} {\bibfnamefont {F.}~\bibnamefont {Ishikawa}}, \ and\ \bibinfo {author} {\bibfnamefont {I.~A.}\ \bibnamefont {Buyanova}},\ }\bibfield  {title} {\enquote {\bibinfo {title} {{Suppression of non-radiative surface recombination by N incorporation in GaAs/GaNAs core/shell nanowires}},}\ }\href {\doibase 10.1038/srep11653} {\bibfield  {journal} {\bibinfo  {journal} {Scientific Reports 2015 5:1}\ }\textbf {\bibinfo {volume} {5}},\ \bibinfo {pages} {1--9} (\bibinfo {year} {2015})}\BibitemShut {NoStop}%
\bibitem [{\citenamefont {Chen}\ \emph {et~al.}(2011)\citenamefont {Chen}, \citenamefont {Tran}, \citenamefont {Ng}, \citenamefont {Ko}, \citenamefont {Chuang}, \citenamefont {Sedgwick},\ and\ \citenamefont {Chang-Hasnain}}]{Chen2011NanolasersSilicon}%
  \BibitemOpen
  \bibfield  {author} {\bibinfo {author} {\bibfnamefont {R.}~\bibnamefont {Chen}}, \bibinfo {author} {\bibfnamefont {T.~T.~D.}\ \bibnamefont {Tran}}, \bibinfo {author} {\bibfnamefont {K.~W.}\ \bibnamefont {Ng}}, \bibinfo {author} {\bibfnamefont {W.~S.}\ \bibnamefont {Ko}}, \bibinfo {author} {\bibfnamefont {L.~C.}\ \bibnamefont {Chuang}}, \bibinfo {author} {\bibfnamefont {F.~G.}\ \bibnamefont {Sedgwick}}, \ and\ \bibinfo {author} {\bibfnamefont {C.}~\bibnamefont {Chang-Hasnain}},\ }\bibfield  {title} {\enquote {\bibinfo {title} {{Nanolasers grown on silicon}},}\ }\href {\doibase 10.1038/nphoton.2010.315} {\bibfield  {journal} {\bibinfo  {journal} {Nature Photonics}\ }\textbf {\bibinfo {volume} {5}},\ \bibinfo {pages} {170--175} (\bibinfo {year} {2011})}\BibitemShut {NoStop}%
\bibitem [{\citenamefont {Paramasivam}, \citenamefont {Gowthaman},\ and\ \citenamefont {Srivastava}(2023)}]{Paramasivam2023Self-consistentHeterostructure}%
  \BibitemOpen
  \bibfield  {author} {\bibinfo {author} {\bibfnamefont {P.}~\bibnamefont {Paramasivam}}, \bibinfo {author} {\bibfnamefont {N.}~\bibnamefont {Gowthaman}}, \ and\ \bibinfo {author} {\bibfnamefont {V.~M.}\ \bibnamefont {Srivastava}},\ }\bibfield  {title} {\enquote {\bibinfo {title} {{Self-consistent Analysis for Optimization of AlGaAs/GaAs Based Heterostructure}},}\ }\href {\doibase 10.1007/S42835-023-01721-7/TABLES/2} {\bibfield  {journal} {\bibinfo  {journal} {Journal of Electrical Engineering and Technology}\ ,\ \bibinfo {pages} {1--15}} (\bibinfo {year} {2023})}\BibitemShut {NoStop}%
\bibitem [{\citenamefont {Lauhon}\ \emph {et~al.}(2002)\citenamefont {Lauhon}, \citenamefont {Gudlksen}, \citenamefont {Wang},\ and\ \citenamefont {Lieber}}]{Lauhon2002EpitaxialHeterostructures}%
  \BibitemOpen
  \bibfield  {author} {\bibinfo {author} {\bibfnamefont {L.~J.}\ \bibnamefont {Lauhon}}, \bibinfo {author} {\bibfnamefont {M.~S.}\ \bibnamefont {Gudlksen}}, \bibinfo {author} {\bibfnamefont {D.}~\bibnamefont {Wang}}, \ and\ \bibinfo {author} {\bibfnamefont {C.~M.}\ \bibnamefont {Lieber}},\ }\bibfield  {title} {\enquote {\bibinfo {title} {{Epitaxial core–shell and core–multishell nanowire heterostructures}},}\ }\href {\doibase 10.1038/nature01141} {\bibfield  {journal} {\bibinfo  {journal} {Nature 2002 420:6911}\ }\textbf {\bibinfo {volume} {420}},\ \bibinfo {pages} {57--61} (\bibinfo {year} {2002})}\BibitemShut {NoStop}%
\bibitem [{\citenamefont {Glas}\ \emph {et~al.}(2013)\citenamefont {Glas}, \citenamefont {Ramdani}, \citenamefont {Patriarche},\ and\ \citenamefont {Harmand}}]{Glas2013PredictiveGrowth}%
  \BibitemOpen
  \bibfield  {author} {\bibinfo {author} {\bibfnamefont {F.}~\bibnamefont {Glas}}, \bibinfo {author} {\bibfnamefont {M.~R.}\ \bibnamefont {Ramdani}}, \bibinfo {author} {\bibfnamefont {G.}~\bibnamefont {Patriarche}}, \ and\ \bibinfo {author} {\bibfnamefont {J.~C.}\ \bibnamefont {Harmand}},\ }\bibfield  {title} {\enquote {\bibinfo {title} {{Predictive modeling of self-catalyzed III-V nanowire growth}},}\ }\href {\doibase 10.1103/PHYSREVB.88.195304/FIGURES/10/MEDIUM} {\bibfield  {journal} {\bibinfo  {journal} {Physical Review B - Condensed Matter and Materials Physics}\ }\textbf {\bibinfo {volume} {88}},\ \bibinfo {pages} {195304} (\bibinfo {year} {2013})}\BibitemShut {NoStop}%
\bibitem [{\citenamefont {Barrig{\'{o}}n}\ \emph {et~al.}(2019)\citenamefont {Barrig{\'{o}}n}, \citenamefont {Heurlin}, \citenamefont {Bi}, \citenamefont {Monemar},\ and\ \citenamefont {Samuelson}}]{Barrigon2019SynthesisNanowires}%
  \BibitemOpen
  \bibfield  {author} {\bibinfo {author} {\bibfnamefont {E.}~\bibnamefont {Barrig{\'{o}}n}}, \bibinfo {author} {\bibfnamefont {M.}~\bibnamefont {Heurlin}}, \bibinfo {author} {\bibfnamefont {Z.}~\bibnamefont {Bi}}, \bibinfo {author} {\bibfnamefont {B.}~\bibnamefont {Monemar}}, \ and\ \bibinfo {author} {\bibfnamefont {L.}~\bibnamefont {Samuelson}},\ }\bibfield  {title} {\enquote {\bibinfo {title} {{Synthesis and Applications of III-V Nanowires}},}\ }\href {\doibase 10.1021/ACS.CHEMREV.9B00075/ASSET/IMAGES/LARGE/CR-2019-00075A{\_}0024.JPEG} {\bibfield  {journal} {\bibinfo  {journal} {Chemical Reviews}\ }\textbf {\bibinfo {volume} {119}},\ \bibinfo {pages} {9170--9220} (\bibinfo {year} {2019})}\BibitemShut {NoStop}%
\bibitem [{\citenamefont {Jabeen}\ \emph {et~al.}(2008)\citenamefont {Jabeen}, \citenamefont {Grillo}, \citenamefont {Rubini},\ and\ \citenamefont {Martelli}}]{Jabeen2008Self-catalyzedEpitaxy}%
  \BibitemOpen
  \bibfield  {author} {\bibinfo {author} {\bibfnamefont {F.}~\bibnamefont {Jabeen}}, \bibinfo {author} {\bibfnamefont {V.}~\bibnamefont {Grillo}}, \bibinfo {author} {\bibfnamefont {S.}~\bibnamefont {Rubini}}, \ and\ \bibinfo {author} {\bibfnamefont {F.}~\bibnamefont {Martelli}},\ }\bibfield  {title} {\enquote {\bibinfo {title} {{Self-catalyzed growth of GaAs nanowires on cleaved Si by molecular beam epitaxy}},}\ }\href {\doibase 10.1088/0957-4484/19/27/275711} {\bibfield  {journal} {\bibinfo  {journal} {Nanotechnology}\ }\textbf {\bibinfo {volume} {19}},\ \bibinfo {pages} {275711} (\bibinfo {year} {2008})}\BibitemShut {NoStop}%
\bibitem [{\citenamefont {Church}\ \emph {et~al.}(2022)\citenamefont {Church}, \citenamefont {Choi}, \citenamefont {Al-Amairi}, \citenamefont {Al-Abri}, \citenamefont {Sanders}, \citenamefont {Oksenberg}, \citenamefont {Joselevich},\ and\ \citenamefont {Parkinson}}]{Church2022HolisticNanowires}%
  \BibitemOpen
  \bibfield  {author} {\bibinfo {author} {\bibfnamefont {S.~A.}\ \bibnamefont {Church}}, \bibinfo {author} {\bibfnamefont {H.}~\bibnamefont {Choi}}, \bibinfo {author} {\bibfnamefont {N.}~\bibnamefont {Al-Amairi}}, \bibinfo {author} {\bibfnamefont {R.}~\bibnamefont {Al-Abri}}, \bibinfo {author} {\bibfnamefont {E.}~\bibnamefont {Sanders}}, \bibinfo {author} {\bibfnamefont {E.}~\bibnamefont {Oksenberg}}, \bibinfo {author} {\bibfnamefont {E.}~\bibnamefont {Joselevich}}, \ and\ \bibinfo {author} {\bibfnamefont {P.~W.}\ \bibnamefont {Parkinson}},\ }\bibfield  {title} {\enquote {\bibinfo {title} {{Holistic Determination of Optoelectronic Properties using High-Throughput Spectroscopy of Surface-Guided CsPbBr3 Nanowires}},}\ }\href {\doibase 10.1021/acsnano.2c01086} {\bibfield  {journal} {\bibinfo  {journal} {ACS Nano}\ }\textbf {\bibinfo {volume} {16}},\ \bibinfo {pages} {9086--9094} (\bibinfo {year} {2022})}\BibitemShut {NoStop}%
\bibitem [{\citenamefont {Jia}\ \emph {et~al.}(2019)\citenamefont {Jia}, \citenamefont {Lin}, \citenamefont {Huang},\ and\ \citenamefont {Duan}}]{Jia2019NanowireMacroscale}%
  \BibitemOpen
  \bibfield  {author} {\bibinfo {author} {\bibfnamefont {C.}~\bibnamefont {Jia}}, \bibinfo {author} {\bibfnamefont {Z.}~\bibnamefont {Lin}}, \bibinfo {author} {\bibfnamefont {Y.}~\bibnamefont {Huang}}, \ and\ \bibinfo {author} {\bibfnamefont {X.}~\bibnamefont {Duan}},\ }\bibfield  {title} {\enquote {\bibinfo {title} {{Nanowire Electronics: From Nanoscale to Macroscale}},}\ }\href {\doibase 10.1021/ACS.CHEMREV.9B00164} {\bibfield  {journal} {\bibinfo  {journal} {Chemical Reviews}\ }\textbf {\bibinfo {volume} {119}},\ \bibinfo {pages} {9074--9135} (\bibinfo {year} {2019})}\BibitemShut {NoStop}%
\bibitem [{\citenamefont {Minehisa}\ \emph {et~al.}(2023)\citenamefont {Minehisa}, \citenamefont {Murakami}, \citenamefont {Hashimoto}, \citenamefont {Nakama}, \citenamefont {Sakaguchi}, \citenamefont {Tsutsumi}, \citenamefont {Tanigawa}, \citenamefont {Yukimune}, \citenamefont {Nagashima}, \citenamefont {Yanagida}, \citenamefont {Sato}, \citenamefont {Hiura}, \citenamefont {Murayama},\ and\ \citenamefont {Ishikawa}}]{Minehisa2023Wafer-scaleEpitaxy}%
  \BibitemOpen
  \bibfield  {author} {\bibinfo {author} {\bibfnamefont {K.}~\bibnamefont {Minehisa}}, \bibinfo {author} {\bibfnamefont {R.}~\bibnamefont {Murakami}}, \bibinfo {author} {\bibfnamefont {H.}~\bibnamefont {Hashimoto}}, \bibinfo {author} {\bibfnamefont {K.}~\bibnamefont {Nakama}}, \bibinfo {author} {\bibfnamefont {K.}~\bibnamefont {Sakaguchi}}, \bibinfo {author} {\bibfnamefont {R.}~\bibnamefont {Tsutsumi}}, \bibinfo {author} {\bibfnamefont {T.}~\bibnamefont {Tanigawa}}, \bibinfo {author} {\bibfnamefont {M.}~\bibnamefont {Yukimune}}, \bibinfo {author} {\bibfnamefont {K.}~\bibnamefont {Nagashima}}, \bibinfo {author} {\bibfnamefont {T.}~\bibnamefont {Yanagida}}, \bibinfo {author} {\bibfnamefont {S.}~\bibnamefont {Sato}}, \bibinfo {author} {\bibfnamefont {S.}~\bibnamefont {Hiura}}, \bibinfo {author} {\bibfnamefont {A.}~\bibnamefont {Murayama}}, \ and\ \bibinfo {author} {\bibfnamefont {F.}~\bibnamefont {Ishikawa}},\ }\bibfield  {title} {\enquote {\bibinfo {title} {{Wafer-scale integration of GaAs/AlGaAs core-shell
  nanowires on silicon by the single process of self-catalyzed molecular beam epitaxy}},}\ }\href {\doibase 10.1039/d2na00848c} {\bibfield  {journal} {\bibinfo  {journal} {Nanoscale Advances}\ }\textbf {\bibinfo {volume} {5}},\ \bibinfo {pages} {1651--1663} (\bibinfo {year} {2023})}\BibitemShut {NoStop}%
\bibitem [{\citenamefont {Peng}\ and\ \citenamefont {Lee}(2011)}]{Peng2011SiliconConversion}%
  \BibitemOpen
  \bibfield  {author} {\bibinfo {author} {\bibfnamefont {K.~Q.}\ \bibnamefont {Peng}}\ and\ \bibinfo {author} {\bibfnamefont {S.~T.}\ \bibnamefont {Lee}},\ }\bibfield  {title} {\enquote {\bibinfo {title} {{Silicon Nanowires for Photovoltaic Solar Energy Conversion}},}\ }\href {\doibase 10.1002/ADMA.201002410} {\bibfield  {journal} {\bibinfo  {journal} {Advanced Materials}\ }\textbf {\bibinfo {volume} {23}},\ \bibinfo {pages} {198--215} (\bibinfo {year} {2011})}\BibitemShut {NoStop}%
\bibitem [{\citenamefont {Amato}\ and\ \citenamefont {Rurali}(2016)}]{Amato2016SurfaceNanowires}%
  \BibitemOpen
  \bibfield  {author} {\bibinfo {author} {\bibfnamefont {M.}~\bibnamefont {Amato}}\ and\ \bibinfo {author} {\bibfnamefont {R.}~\bibnamefont {Rurali}},\ }\bibfield  {title} {\enquote {\bibinfo {title} {{Surface physics of semiconducting nanowires}},}\ }\href {\doibase 10.1016/J.PROGSURF.2015.11.001} {\bibfield  {journal} {\bibinfo  {journal} {Progress in Surface Science}\ }\textbf {\bibinfo {volume} {91}},\ \bibinfo {pages} {1--28} (\bibinfo {year} {2016})}\BibitemShut {NoStop}%
\bibitem [{\citenamefont {Fonseka}\ \emph {et~al.}(2019)\citenamefont {Fonseka}, \citenamefont {Caroff}, \citenamefont {Guo}, \citenamefont {Sanchez}, \citenamefont {Tan},\ and\ \citenamefont {Jagadish}}]{Fonseka2019EngineeringHeterostructures}%
  \BibitemOpen
  \bibfield  {author} {\bibinfo {author} {\bibfnamefont {H.~A.}\ \bibnamefont {Fonseka}}, \bibinfo {author} {\bibfnamefont {P.}~\bibnamefont {Caroff}}, \bibinfo {author} {\bibfnamefont {Y.}~\bibnamefont {Guo}}, \bibinfo {author} {\bibfnamefont {A.~M.}\ \bibnamefont {Sanchez}}, \bibinfo {author} {\bibfnamefont {H.~H.}\ \bibnamefont {Tan}}, \ and\ \bibinfo {author} {\bibfnamefont {C.}~\bibnamefont {Jagadish}},\ }\bibfield  {title} {\enquote {\bibinfo {title} {{Engineering the Side Facets of Vertical [100] Oriented InP Nanowires for Novel Radial Heterostructures}},}\ }\href {\doibase 10.1186/s11671-019-3177-6} {\bibfield  {journal} {\bibinfo  {journal} {Nanoscale Research Letters}\ }\textbf {\bibinfo {volume} {14}} (\bibinfo {year} {2019}),\ 10.1186/s11671-019-3177-6}\BibitemShut {NoStop}%
\bibitem [{\citenamefont {Fedorov}\ \emph {et~al.}(2021)\citenamefont {Fedorov}, \citenamefont {Berdnikov}, \citenamefont {Sibirev}, \citenamefont {Bolshakov}, \citenamefont {Fedina}, \citenamefont {Sapunov}, \citenamefont {Dvoretckaia}, \citenamefont {Cirlin}, \citenamefont {Kirilenko}, \citenamefont {Tchernycheva},\ and\ \citenamefont {Mukhin}}]{Fedorov2021TailoringSubstrates}%
  \BibitemOpen
  \bibfield  {author} {\bibinfo {author} {\bibfnamefont {V.~V.}\ \bibnamefont {Fedorov}}, \bibinfo {author} {\bibfnamefont {Y.}~\bibnamefont {Berdnikov}}, \bibinfo {author} {\bibfnamefont {N.~V.}\ \bibnamefont {Sibirev}}, \bibinfo {author} {\bibfnamefont {A.~D.}\ \bibnamefont {Bolshakov}}, \bibinfo {author} {\bibfnamefont {S.~V.}\ \bibnamefont {Fedina}}, \bibinfo {author} {\bibfnamefont {G.~A.}\ \bibnamefont {Sapunov}}, \bibinfo {author} {\bibfnamefont {L.~N.}\ \bibnamefont {Dvoretckaia}}, \bibinfo {author} {\bibfnamefont {G.}~\bibnamefont {Cirlin}}, \bibinfo {author} {\bibfnamefont {D.~A.}\ \bibnamefont {Kirilenko}}, \bibinfo {author} {\bibfnamefont {M.}~\bibnamefont {Tchernycheva}}, \ and\ \bibinfo {author} {\bibfnamefont {I.~S.}\ \bibnamefont {Mukhin}},\ }\bibfield  {title} {\enquote {\bibinfo {title} {{Tailoring morphology and vertical yield of self-catalyzed gap nanowires on template-free si substrates}},}\ }\href {\doibase 10.3390/NANO11081949/S1} {\bibfield  {journal} {\bibinfo  {journal}
  {Nanomaterials}\ }\textbf {\bibinfo {volume} {11}},\ \bibinfo {pages} {1949} (\bibinfo {year} {2021})}\BibitemShut {NoStop}%
\bibitem [{\citenamefont {Aspnes}\ and\ \citenamefont {Studna}(1983)}]{Aspnes1983DielectricEV}%
  \BibitemOpen
  \bibfield  {author} {\bibinfo {author} {\bibfnamefont {D.~E.}\ \bibnamefont {Aspnes}}\ and\ \bibinfo {author} {\bibfnamefont {A.~A.}\ \bibnamefont {Studna}},\ }\bibfield  {title} {\enquote {\bibinfo {title} {{Dielectric functions and optical parameters of Si, Ge, GaP, GaAs, GaSb, InP, InAs, and InSb from 1.5 to 6.0 eV}},}\ }\href {\doibase 10.1103/PhysRevB.27.985} {\bibfield  {journal} {\bibinfo  {journal} {Physical Review B}\ }\textbf {\bibinfo {volume} {27}},\ \bibinfo {pages} {985} (\bibinfo {year} {1983})}\BibitemShut {NoStop}%
\bibitem [{\citenamefont {Pevere}\ \emph {et~al.}(2018)\citenamefont {Pevere}, \citenamefont {Sangghaleh}, \citenamefont {Bruhn}, \citenamefont {Sychugov},\ and\ \citenamefont {Linnros}}]{Pevere2018RapidNanocrystals}%
  \BibitemOpen
  \bibfield  {author} {\bibinfo {author} {\bibfnamefont {F.}~\bibnamefont {Pevere}}, \bibinfo {author} {\bibfnamefont {F.}~\bibnamefont {Sangghaleh}}, \bibinfo {author} {\bibfnamefont {B.}~\bibnamefont {Bruhn}}, \bibinfo {author} {\bibfnamefont {I.}~\bibnamefont {Sychugov}}, \ and\ \bibinfo {author} {\bibfnamefont {J.}~\bibnamefont {Linnros}},\ }\bibfield  {title} {\enquote {\bibinfo {title} {{Rapid Trapping as the Origin of Nonradiative Recombination in Semiconductor Nanocrystals}},}\ }\href {\doibase 10.1021/ACSPHOTONICS.8B00581/ASSET/IMAGES/LARGE/PH-2018-00581C{\_}0005.JPEG} {\bibfield  {journal} {\bibinfo  {journal} {ACS Photonics}\ }\textbf {\bibinfo {volume} {5}},\ \bibinfo {pages} {2990--2996} (\bibinfo {year} {2018})}\BibitemShut {NoStop}%
\bibitem [{\citenamefont {Niemeyer}\ \emph {et~al.}(2019)\citenamefont {Niemeyer}, \citenamefont {Kleinschmidt}, \citenamefont {Walker}, \citenamefont {Mundt}, \citenamefont {Timm}, \citenamefont {Lang}, \citenamefont {Hannappel},\ and\ \citenamefont {Lackner}}]{Niemeyer2019MeasurementGaAs}%
  \BibitemOpen
  \bibfield  {author} {\bibinfo {author} {\bibfnamefont {M.}~\bibnamefont {Niemeyer}}, \bibinfo {author} {\bibfnamefont {P.}~\bibnamefont {Kleinschmidt}}, \bibinfo {author} {\bibfnamefont {A.~W.}\ \bibnamefont {Walker}}, \bibinfo {author} {\bibfnamefont {L.~E.}\ \bibnamefont {Mundt}}, \bibinfo {author} {\bibfnamefont {C.}~\bibnamefont {Timm}}, \bibinfo {author} {\bibfnamefont {R.}~\bibnamefont {Lang}}, \bibinfo {author} {\bibfnamefont {T.}~\bibnamefont {Hannappel}}, \ and\ \bibinfo {author} {\bibfnamefont {D.}~\bibnamefont {Lackner}},\ }\bibfield  {title} {\enquote {\bibinfo {title} {{Measurement of the non-radiative minority recombination lifetime and the effective radiative recombination coefficient in GaAs}},}\ }\href {\doibase 10.1063/1.5051709/1076249} {\bibfield  {journal} {\bibinfo  {journal} {AIP Advances}\ }\textbf {\bibinfo {volume} {9}},\ \bibinfo {pages} {45034} (\bibinfo {year} {2019})}\BibitemShut {NoStop}%
\bibitem [{\citenamefont {Furthmeier}\ \emph {et~al.}(2014)\citenamefont {Furthmeier}, \citenamefont {Dirnberger}, \citenamefont {Hubmann}, \citenamefont {Bauer}, \citenamefont {Korn}, \citenamefont {Sch{\"{u}}ller}, \citenamefont {Zweck}, \citenamefont {Reiger},\ and\ \citenamefont {Bougeard}}]{Furthmeier2014LongNanowires}%
  \BibitemOpen
  \bibfield  {author} {\bibinfo {author} {\bibfnamefont {S.}~\bibnamefont {Furthmeier}}, \bibinfo {author} {\bibfnamefont {F.}~\bibnamefont {Dirnberger}}, \bibinfo {author} {\bibfnamefont {J.}~\bibnamefont {Hubmann}}, \bibinfo {author} {\bibfnamefont {B.}~\bibnamefont {Bauer}}, \bibinfo {author} {\bibfnamefont {T.}~\bibnamefont {Korn}}, \bibinfo {author} {\bibfnamefont {C.}~\bibnamefont {Sch{\"{u}}ller}}, \bibinfo {author} {\bibfnamefont {J.}~\bibnamefont {Zweck}}, \bibinfo {author} {\bibfnamefont {E.}~\bibnamefont {Reiger}}, \ and\ \bibinfo {author} {\bibfnamefont {D.}~\bibnamefont {Bougeard}},\ }\bibfield  {title} {\enquote {\bibinfo {title} {{Long exciton lifetimes in stacking-fault-free wurtzite GaAs nanowires}},}\ }\href {\doibase 10.1063/1.4903482/132901} {\bibfield  {journal} {\bibinfo  {journal} {Applied Physics Letters}\ }\textbf {\bibinfo {volume} {105}},\ \bibinfo {pages} {222109} (\bibinfo {year} {2014})}\BibitemShut {NoStop}%
\bibitem [{\citenamefont {Joyce}\ \emph {et~al.}(2008)\citenamefont {Joyce}, \citenamefont {Gao}, \citenamefont {Tan}, \citenamefont {Jagadish}, \citenamefont {Kim}, \citenamefont {Fickenscher}, \citenamefont {Perera}, \citenamefont {Hoang}, \citenamefont {Smith}, \citenamefont {Jackson}, \citenamefont {Yarrison-Rice}, \citenamefont {Zhang},\ and\ \citenamefont {Zou}}]{Joyce2008HighCharacterization}%
  \BibitemOpen
  \bibfield  {author} {\bibinfo {author} {\bibfnamefont {H.~J.}\ \bibnamefont {Joyce}}, \bibinfo {author} {\bibfnamefont {Q.}~\bibnamefont {Gao}}, \bibinfo {author} {\bibfnamefont {H.~H.}\ \bibnamefont {Tan}}, \bibinfo {author} {\bibfnamefont {C.}~\bibnamefont {Jagadish}}, \bibinfo {author} {\bibfnamefont {Y.}~\bibnamefont {Kim}}, \bibinfo {author} {\bibfnamefont {M.~A.}\ \bibnamefont {Fickenscher}}, \bibinfo {author} {\bibfnamefont {S.}~\bibnamefont {Perera}}, \bibinfo {author} {\bibfnamefont {T.~B.}\ \bibnamefont {Hoang}}, \bibinfo {author} {\bibfnamefont {L.~M.}\ \bibnamefont {Smith}}, \bibinfo {author} {\bibfnamefont {H.~E.}\ \bibnamefont {Jackson}}, \bibinfo {author} {\bibfnamefont {J.~M.}\ \bibnamefont {Yarrison-Rice}}, \bibinfo {author} {\bibfnamefont {X.}~\bibnamefont {Zhang}}, \ and\ \bibinfo {author} {\bibfnamefont {J.}~\bibnamefont {Zou}},\ }\bibfield  {title} {\enquote {\bibinfo {title} {{High purity GaAs nanowires free of planar defects: Growth and characterization}},}\ }\href {\doibase
  10.1002/ADFM.200800625} {\bibfield  {journal} {\bibinfo  {journal} {Advanced Functional Materials}\ }\textbf {\bibinfo {volume} {18}},\ \bibinfo {pages} {3794--3800} (\bibinfo {year} {2008})}\BibitemShut {NoStop}%
\bibitem [{\citenamefont {Plissard}\ \emph {et~al.}(2011)\citenamefont {Plissard}, \citenamefont {Larrieu}, \citenamefont {Wallart},\ and\ \citenamefont {Caroff}}]{Plissard2011HighDropletpositioning}%
  \BibitemOpen
  \bibfield  {author} {\bibinfo {author} {\bibfnamefont {S.}~\bibnamefont {Plissard}}, \bibinfo {author} {\bibfnamefont {G.}~\bibnamefont {Larrieu}}, \bibinfo {author} {\bibfnamefont {X.}~\bibnamefont {Wallart}}, \ and\ \bibinfo {author} {\bibfnamefont {P.}~\bibnamefont {Caroff}},\ }\bibfield  {title} {\enquote {\bibinfo {title} {{High yield of self-catalyzed GaAs nanowire arrays grown on silicon via gallium dropletpositioning}},}\ }\href {\doibase 10.1088/0957-4484/22/27/275602} {\bibfield  {journal} {\bibinfo  {journal} {Nanotechnology}\ }\textbf {\bibinfo {volume} {22}},\ \bibinfo {pages} {275602} (\bibinfo {year} {2011})}\BibitemShut {NoStop}%
\bibitem [{\citenamefont {Kim}\ \emph {et~al.}()\citenamefont {Kim}, \citenamefont {Joyce}, \citenamefont {Gao}, \citenamefont {Hoe~Tan}, \citenamefont {Jagadish}, \citenamefont {Paladugu}, \citenamefont {Zou},\ and\ \citenamefont {Suvorova}}]{KimInfluenceNanowires}%
  \BibitemOpen
  \bibfield  {author} {\bibinfo {author} {\bibfnamefont {Y.}~\bibnamefont {Kim}}, \bibinfo {author} {\bibfnamefont {H.~J.}\ \bibnamefont {Joyce}}, \bibinfo {author} {\bibfnamefont {Q.}~\bibnamefont {Gao}}, \bibinfo {author} {\bibfnamefont {H.}~\bibnamefont {Hoe~Tan}}, \bibinfo {author} {\bibfnamefont {C.}~\bibnamefont {Jagadish}}, \bibinfo {author} {\bibfnamefont {M.}~\bibnamefont {Paladugu}}, \bibinfo {author} {\bibfnamefont {J.}~\bibnamefont {Zou}}, \ and\ \bibinfo {author} {\bibfnamefont {A.~A.}\ \bibnamefont {Suvorova}},\ }\bibfield  {title} {\enquote {\bibinfo {title} {{Influence of Nanowire Density on the Shape and Optical Properties of Ternary InGaAs Nanowires}},}\ }\href {\doibase 10.1021/nl052189o} {\ 10.1021/nl052189o}\BibitemShut {NoStop}%
\bibitem [{\citenamefont {Casey}\ \emph {et~al.}(1975)\citenamefont {Casey}, \citenamefont {Sell}, \citenamefont {Wecht}, \citenamefont {Phys~Lett},\ and\ \citenamefont {Casey~Jr}}]{Casey1975ConcentrationEV}%
  \BibitemOpen
  \bibfield  {author} {\bibinfo {author} {\bibfnamefont {H.~C.}\ \bibnamefont {Casey}}, \bibinfo {author} {\bibfnamefont {D.~D.}\ \bibnamefont {Sell}}, \bibinfo {author} {\bibfnamefont {K.~W.}\ \bibnamefont {Wecht}}, \bibinfo {author} {\bibfnamefont {A.}~\bibnamefont {Phys~Lett}}, \ and\ \bibinfo {author} {\bibfnamefont {H.~C.}\ \bibnamefont {Casey~Jr}},\ }\bibfield  {title} {\enquote {\bibinfo {title} {{Concentration dependence of the absorption coefficient for n and p type GaAs between 1.3 and 1.6 eV}},}\ }\href {\doibase 10.1063/1.321330} {\bibfield  {journal} {\bibinfo  {journal} {J. Appl. Phys}\ }\textbf {\bibinfo {volume} {46}},\ \bibinfo {pages} {5} (\bibinfo {year}
  {1975})}\BibitemShut {NoStop}%
\end{thebibliography}

\end{document}